%% file: main.tex
\documentclass[a4-paper]{article}
\usepackage[utf8]{inputenc}

\usepackage{biblatex}
\usepackage{graphicx}
\usepackage[absolute]{textpos}
\usepackage{zref-user}
\usepackage[dvipsnames]{xcolor}
\usepackage{zref-abspos}
\usepackage{calc}
\usepackage{amsmath}
\usepackage{float}
\usepackage{subcaption}
\usepackage{caption}
\usepackage{multirow}
\usepackage{listings}
\usepackage[normalem]{ulem}
\usepackage[margin=1in]{geometry}
\usepackage{enumitem}
\usepackage[font=small,labelfont=bf]{caption}
\usepackage{gensymb}
\usepackage{hyperref}
\usepackage{tabularx}

\usepackage{mathtools} 

\hypersetup{
    colorlinks,
    citecolor=black,
    filecolor=black,
    linkcolor=black,
    urlcolor=black
}
\usepackage{color} 
\definecolor{mygreen}{RGB}{0,128,0} 
\definecolor{mylilas}{RGB}{170,55,241}
\usepackage{titlesec}

\begin{document}

\begin{titlepage}
\title{
	\begin{center}
	\begin{figure}[h!]
		\centering
		\includegraphics[width=0.5\textwidth]{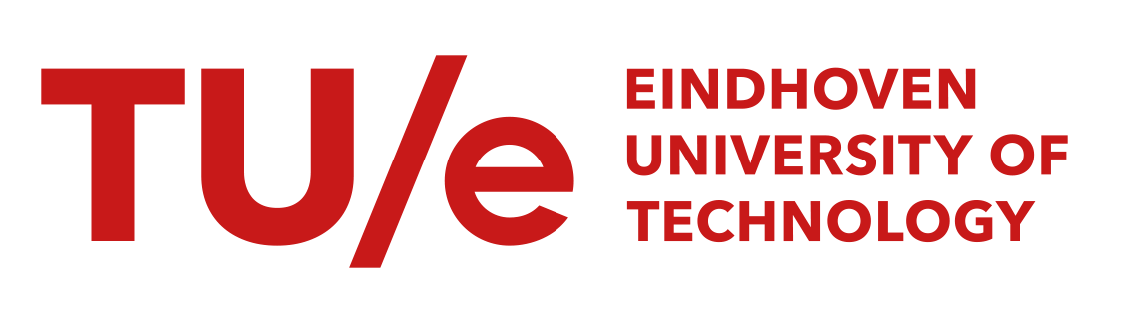}
	\end{figure}
    \Huge{\bfseries Collisional radiative modelling with improved cross sections to investigate plasma molecular interactions in divertor plasmas.} \\
    \line(1,0){400} \\[1cm]
    Internship report
    \end{center}
}

\author{Author: Stijn Kobussen, 1341596 \\ \\ TU/e academic advisor: Dr. Hugo de Blank \\ External UKAEA supervisor: Dr. Kevin Verhaegh}

\date{July 11, 2023}
\maketitle

\begin{figure}[h]
    \includegraphics[width=0.3\textwidth]{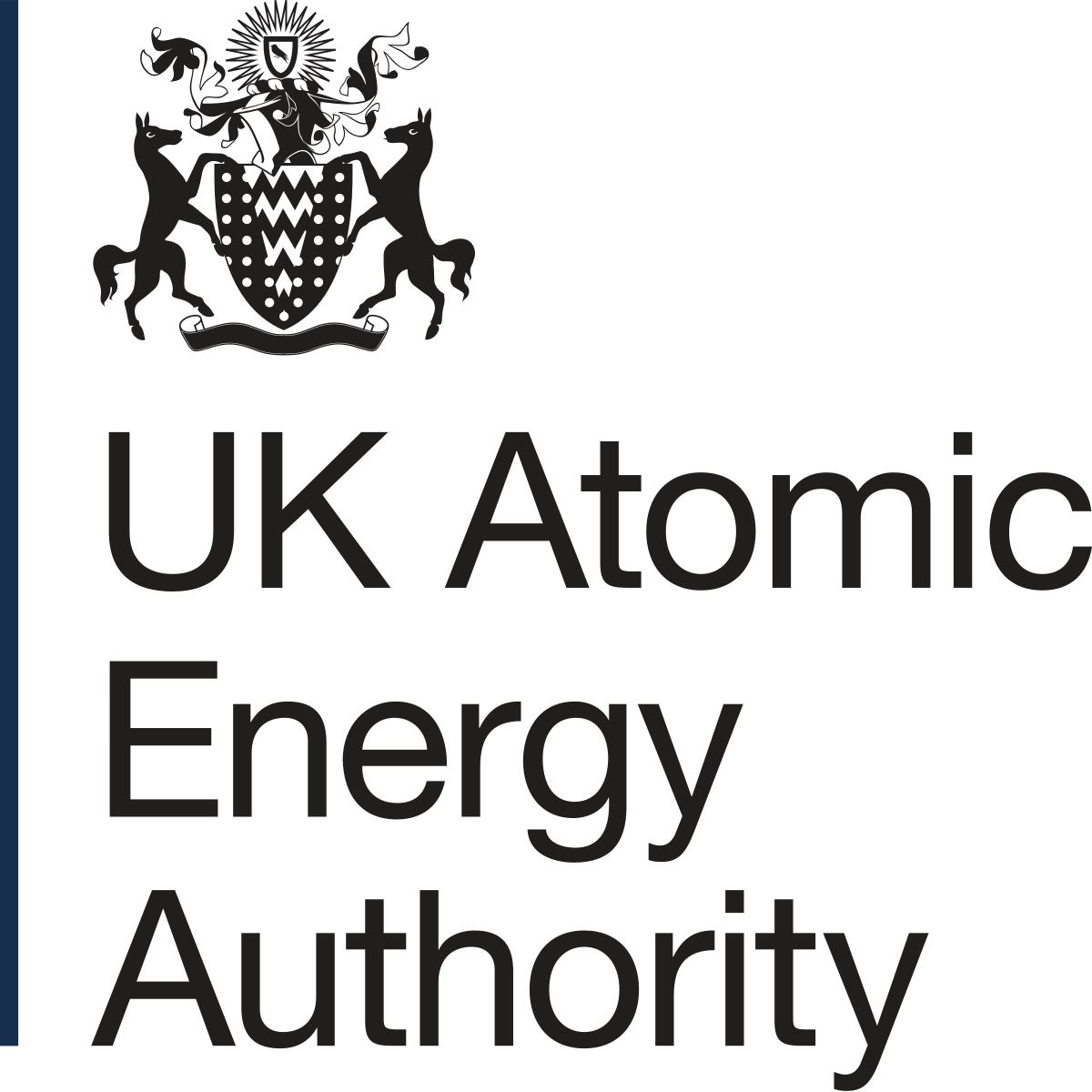}
\end{figure}
\end{titlepage}

    
    
    

\lstset{language=Matlab,%
    breaklines=true,%
    morekeywords={matlab2tikz},
    keywordstyle=\color{blue},%
    morekeywords=[2]{1}, keywordstyle=[2]{\color{black}},
    identifierstyle=\color{black},%
    stringstyle=\color{mylilas},
    commentstyle=\color{mygreen},%
    showstringspaces=false,
    numbers=left,%
    numberstyle={\tiny \color{black}},
    numbersep=9pt, 
    emph=[1]{for,end,break},emphstyle=[1]\color{red}, 
}
\begin{abstract}

\input{Abstract.tex}

\end{abstract}

\input{1_Introduction.tex}

\input{2_Background.tex}

\input{3_Results.tex}
\input{4_Conclusions.tex}

\input{5_Discussion.tex}

\input{Acknowledgements.tex}

\newpage
\appendix

\input{Appendix_A.tex}

\input{Appendix_B.tex}
\input{Appendix_C.tex}

\printbibliography
\end{document}

%% file: Abstract.tex
Power exhaust is one of the main challenges for the realization of practical fusion energy production. The magnetic confinement approach to fusion often uses a divertor configuration, where power loads are critical. Recent SOLPS simulations of divertor plasmas in MAST-U and TCV show significant deviations from the experimental results \cite{verhaegh2023role,Verhaegh_2021}. A possible explanation for this is the incorrect incorporation of plasma-molecular interactions in current SOLPS simulations. SOLPS uses tabulated rate data for the particle physics part of the simulations \cite{WIESEN2015480, Eirene}. There are, however, large concerns about the accuracy of these rates, especially for plasma-molecular interactions \cite{Verhaegh_2023_SOLPS}. Therefore, in this work, improved rate data from different sources, such as the Laporta\cite{Laporta_2021} and MCCCDB\cite{MCCCDB_exc_exc,MCCCDB_exc_X1,MCCCDB_Ionization} databases was used, to construct a Collisional Radiative Model (CRM) to investigate the plasma molecular interactions in conditions relevant to the MAST-U Super-X divertor. It was shown that using these rates, a better correspondence with experiment is achieved than with the tabulated rates. In particular, increased Molecular Activated Recombination (MAR) and Dissociation (MAD) was observed. Additionally it was found that molecular processes may be a significant contributor to divertor physics. The effect of different processes on the vibrational distribution was also investigated, as well as the radiative emission spectra generated from the CRM. It was demonstrated that re-evaluation of reaction rates for plasma-molecular interactions in plasma-edge modelling is necessary.

%% file: 1_Introduction.tex
\section{Introduction}
\subsection{Overview of the subject}
Power exhaust remains one of the main challenges for the realization of fusion energy production. Practical fusion reactors will need to drastically reduce the ion and heat target fluxes to prevent damage  to the divertor target. One way to simultaneously mitigate heat and particle flux to the target is plasma detachment. 

Plasma detachment happens when the ionization front of the plasma moves away from the target in the divertor. This creates a region where the neutral atom and molecular density is increased. In this region, plasma-molecule interactions can occur that result in simultaneous power, momentum and particle losses in the divertor. These reactions result in excited atoms and rovibronically\footnote{Rotationally, vibrationally and electronically} excited molecules, which emit photons. It is therefore possible to analyse the processes and conditions (density, temperature, impurity content, ...) in this region with spectroscopic techniques. 

For the demonstration power plant DEMO, ITER's successor, it is expected that 95\% of the energy generated by the reactor will have to be dissipated radiatively to prevent melting of the wall material \cite{Fil_2022,EUROfusion_2023}. Such high radiation values require significant plasma detachment. It is unclear whether conventional divertors will be able to achieve this high level of detachment, while still achieving acceptable levels of core performance required for a reactor \cite{Reimerdes_2020}. This problem led to the development of Alternative Divertor Configurations (ADC's), such as the Super-X divertor in MAST-U, which is designed to reduce impurity seeding thresholds for detachment \cite{Fil_2022}. These ADC's use alternative (magnetic) geometries to reduce the heat and particle fluxes to the target. 

A reactor using an ADC can operate with high levels of detachment. Simulations of the MAST-U Super-X divertor and the TCV conventional divertor show deviations from the experimental results during deep detachment \cite{verhaegh2023role,Verhaegh_2021}. One possible explanation could be inaccurate rate data used for molecular reactions, in particular molecular charge exchange \cite{Verhaegh_2023_SOLPS} ($D_2 + D^+ \rightarrow D_2^+ + D$). Molecular charge exchange is expected to drive much of the molecular-activated plasma chemistry in the divertor, resulting in significant particle losses due to Molecular Activated Recombination (MAR) and hydrogenic power losses due to Molecular Activated Dissociation (MAD) \cite{Verhaegh_2023_SOLPS}. Most plasma edge models use tabulated polynomial fits representing rates from various sources, packaged with a code called Eirene \cite{Eirene}, for plasma neutral interactions \cite{Verhaegh_2023_SOLPS}. For simplicity, I refer to these tabulated rates as 'Eirene rates'. Eirene is a multi-purpose Bolzmann-equation Monte Carlo solver commonly used in conjunction with a plasma fluid code to model fusion plasmas \cite{Eirene}. However, there are large concerns about the accuracy of the Eirene effective molecular reaction rates used in models such as SOLPS \cite{Verhaegh_2023_SOLPS,WIESEN2015480}. Eirene rates are typically based on rescaled rates for the vibrational ground state \cite{Greenland2001}, which can deviate significantly from full ab initio quantum mechanical rate calculations to get the full vibrational dependency, particularly for molecular charge exchange \cite{Ichihara_2000}. 

Comparisons between power exhaust experiments in current conventional and alternative divertors and plasma-edge modelling is required to reduce the uncertainty in extrapolating to target heat flux estimates in future reactors. To achieve this, it is important to accurately include molecular interactions in simulations. Whether such molecular interactions could play a key role in reactors is still debated \cite{verhaegh2023role,Verhaegh_2021,Verhaegh_2023_SOLPS,Kukushkin2017}. However, they would likely be more important for reactors that operate in deeply detached conditions and/or have elevated molecular densities (e.g. through neutral baffling), such as ADC's. Much contemporary research is focused on more accurately including these processes in models such as SOLPS, the ultimate goal being to develop new sophisticated plasma modelling tools that work even for the low-temperature, high neutral density conditions of alternative divertors. 

Collisional Radiative Modelling (CRM) is a widely used technique for calculating excited state distributions, which can be used to calculate effective reaction rates. A CRM is a 0D model that takes both collisional and radiative processes into account to obtain the density of different species as a function of time. CRM's are often used to calculate effective rate coefficients that are consequently used as input data for larger plasma edge models such as SOLPS. The technique is especially useful, because it is much less computationally expensive than tracking every excited state individually in plasma edge codes. It is therefore more capable of incorporating large numbers of different quantum states and species.

CRUMPET is a python tool designed for the creation and evaluation of collisional radiative models \cite{HOLM2021100982}. In this report, CRUMPET is used to investigate molecular processes in plasma conditions relevant to the MAST-U Super-X divertor plasma. Reaction cross sections and decay coefficients obtained from advanced quantum mechanical calculations are used, to ensure high accuracy of the results. Using this model, it is possible to computationally analyse molecular interactions in a wide range of plasma conditions, and compare them to results obtained from default Eirene data.

\vspace{5mm}

\subsection{This report}

The main goal of this work is to use CRUMPET in combination with accurate vibrationally resolved rate data for the various reactions to construct a CRM and use it to investigate the effects of molecular interactions in deuterium plasmas with conditions applicable to the MAST-U Super-X divertor. To this end, three different topics have been chosen for this report, each with a number of research questions:

\vspace{5mm}


\textbf{1. What does the new CRM say about the impact of plasma-molecular interactions in exhaust processes?} 

\vspace{5mm}

There are concerns about the rates used by Eirene. The CRM used in this report uses improved rate data gathered from various sources to calculate vibrational distributions and re-evaluate the effective reaction rates. The following research questions were chosen for this subject. 

\emph{
\begin{itemize}
    \item[1.1] How do effective rates (calculated using the new CRM) compare to default rates tabulated in Eirene, and how significant would the rate differences be for the effective ion sources/sinks in the plasma?
    \item[1.2] What is the role of electronically excited molecules in the plasma?
    \item[1.3] What is the relative importance of radiative power dissipation caused by electronically excited molecules compared to excited atoms?
\end{itemize}
}

The calculation of effective rates and comparison to rates from the Eirene documentation AMJUEL is given in \autoref{sec:rates}. The effects of electronically excited molecules are investigated in \autoref{sec:exc}. Lastly, the relative importance of radiative power dissipation caused by electronically excited molecules compared to excited atoms is investigated in \autoref{sec:radiative_loss}. 

\vspace{5mm}

\textbf{2. Can the new CRM be used to improve experimental data analysis of plasma-molecular interactions?}

\vspace{5mm}

Visible spectroscopy is commonly used as a diagnostic tool in plasma physics, and the Fulcher band emission is often used to study the rovibronic states of molecular deuterium in fusion plasmas. Analysis of the Fulcher band spectrum can be used to determine the vibrational distribution in the upper state. By default, this is analysed using a simple model involving Franck-Condon factors \cite{osborne2023initial}. For this reseach subject, the distribution in the upper Fulcher state is investigated to address the following research questions:

\emph{
\begin{itemize}
    \item[2.1] Under which conditions is the simple model involving Franck-Condon factors valid?
    \item[2.2] Can the CRM of this report be used to generate emission spectra that can be used to improve analysis of experimental data?
\end{itemize}
}

By modelling the upper state of the Fulcher band using a fully vibrationally resolved CRM, we can investigate the validity of the Franck condon model. It could also be possible to use the new CRM to generate emission spectra from the Fulcher band. These can be compared to experiment and could possibly be used to obtain information about the molecules, such as density and radiative power emission. The investigation into research questions 2.1 and 2.2 is given in \autoref{sec:Fulcher}.

\vspace{5mm}

\textbf{3. Under which conditions can effective rates for plasma-molecular interactions be properly applied?}

\vspace{5mm}

There are several processes in divertor plasmas which can influence the vibrational distribution in the plasma, that cannot fully be taken into account by a CRM. The conditions under which these processes significantly influence exhaust physics needs to be investigated, as this prevents using effective rates in plasma exhaust codes. Such processes include vibrational-vibrational exchange (VVE), transport effects and plasma-wall interactions (PWI). This gives rise to the following research questions:

\emph{
\begin{itemize}
    \item[3.1] Under which conditions is there a significant effect of vibrational-vibrational exchange on the vibrational distribution; and under which conditions can this significantly impact plasma-molecular reaction rates?
    \item[3.2] Under which conditions can transport have a significant impact on the vibrational distribution of the molecules as well as on the MAR ion sink?
    \item[3.3] Under which conditions can the influx of vibrationally excited molecules from the wall after plasma wall interactions have a significant impact on the MAR ion sink?
\end{itemize} 
}

Collisions between molecules can alter the vibrational distribution through VVE. The importance of VVE is investigated in \autoref{sec:VVE}. Transport inside the divertor may influence the vibrational distribution of the molecules, the effect of which is investigated in \autoref{sec:transport}. PWI can cause an influx of vibrationally excited molecules into the plasma, which could affect the vibrational distribution. The effect of PWI on effective rates is investigated in \autoref{sec:PWI}.

%% file: 2_Background.tex
\section{Background}
\label{sec:background}
This section contains the necessary background information for this report. 

\subsection{Energy levels of molecular hydrogen (and deuterium)}
\label{sec:E_levels}
In this report, interactions between the plasma and molecular deuterium are investigated. These interactions depend on the energy levels of the molecules. Diatomic molecules can be excited electronically, vibrationally, and rotationally, or 'rovibronically' in short. The rovibronic structure of diatomic deuterium is very similar to that of diatomic hydrogen, only the energy levels are shifted slightly \cite{Fantz2002}. 

\begin{figure}[h]
    \centering
    \includegraphics[width=0.5\textwidth]{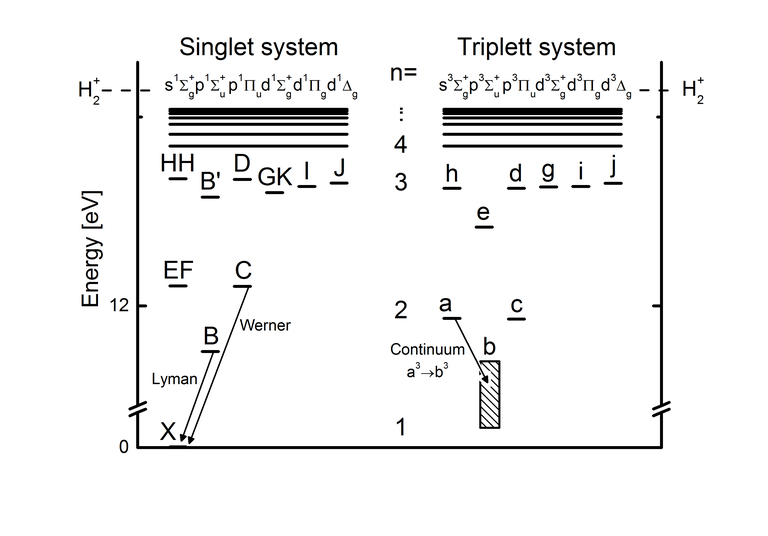}
    \caption{Energy diagram of the electronic levels of molecular hydrogen. Image adapted from the Yacora H$_2$ model help page.}
    \label{fig:E_diagram}
\end{figure}

Electronic excitation corresponds to the transition of electrons from a bound state, to a higher energetic bound state. Because molecular hydrogen consists of two atoms and two electrons, the atomic orbitals of the electrons overlap. This causes many different possible energy states, such as so-called singlet and triplet states. \autoref{fig:E_diagram} contains an energy level diagram of the electronic states of molecular hydrogen. Molecules in the ground state 'X' can be excited to any of the other singlet of triplet states, either by collisions with electrons or absorption of a photon. Once in an excited state, the molecules can decay back to a lower state, be further excited to an even higher state, or react with the plasma in a number of different ways. 

Each of the electronic states in \autoref{fig:E_diagram} is also subdivided into different vibrational states, each with a slightly different energy. The vibrational level of a molecule corresponds to the energy contained in the vibration of the two hydrogen atoms. The vibrational state of the molecules not only affects the energy, but also the molecular reaction rates are highly dependent on the vibrational state \cite{Ichihara_2000}. It is therefore important to take these states into account for the rates used in plasma edge modelling codes, when molecules are expected to play a major role in the plasma chemistry. 

Each vibrational state in each electronic level is further subdivided into rotational levels. The rotational levels are caused by the fact that the atoms in a diatomic molecule can rotate around each other, associated with some rotational energy. These rotational states do not have a significant effect on the reaction rates in the plasma, so they are not included in the model for this report \cite{Sawada1995}. The rotational states do affect the emission spectra from the plasma, so they are important for diagnostics using spectroscopic techniques.

\subsection{Collisional Radiative Models (CRM's)}
\label{sec:CRMs}
This report uses the python tool CRUMPET to generate CRMs to investigate the molecular processes in a divertor tokamak plasma. To provide some context, a short description of how CRMs work is given here. CRUMPET is based on the Greenland CRM equation \cite{HOLM2021100982}. The time-evolution of neutral species interacting with a background plasma is described by a set of nonlinear equations 

\begin{equation}
    \dot{n}_k = \sum_{i,j}R^k_{i,j}(T)n_in_j+\sum_jA^k_jn_j-n_k\sum_{i,j}R^j_{i,k}(T)n_i-n_k\sum A^j_k +\Gamma_k
    \label{eq:Rate_eq}
\end{equation}

where $n_i$ is the density of species $i$, and $R^k_{i,j}$ is the rate coefficient for production of species $k$ due to collisions of $i$ and $j$. $A^k_j$ is the Einstein coefficient for radiative decay from species $j$ to $k$. The first two terms in equation \eqref{eq:Rate_eq} are for collisional and radiative production of species $k$, respectively. The second and third terms are for losses of species $k$ \cite{Greenland_Molecules}. The final term in \eqref{eq:Rate_eq} is the source term of species $k$. This equation is difficult to solve due to its high nonlinearity. The equation can be linearized, however, by fixing the densities of the so-called 'background species', e.g. ions and electrons. If we only consider collisions between background and other species, equation \eqref{eq:Rate_eq} can be linearized and written as 

\begin{equation}
    \dot{\textbf{n}} = \textbf{M}(T,n_i,...,n_m)\textbf{n}+\pmb{\Gamma}(T,n_i,...n_m).
    \label{eq:matrices}
\end{equation}

The densities of all the background species are given by $n_i,...,n_m$, while $\textbf{n}$ is the vector containing all the other species. For a system containing $N$ non-background species, $M$ is an $N\times N$ matrix of rates and depends only on the temperatures of the background species and their densities. $\pmb{\Gamma}$ is a vector containing sinks and sources caused either by collisions between background species, or external sources such as transport. Since $\textbf{M}$ and $\pmb{\Gamma}$ are independent of $\mathbf{n}$ the system is linear \cite{Greenland_Molecules}.

CRUMPET builds a CRM object by loading an input file. This specifies: 1) the relevant (tracked) species, background species and their energies; 2) reaction equations, both collisional and radiative, with a reference to the relative rate data (generally expressed in terms of tables containing a polynomial fit of the rates). 

Given an electron density, electron temperature, this CRM object can construct the matrices of equation \eqref{eq:matrices}. CRUMPET then uses the matrices and standard ODE solvers to obtain either the densities of the (tracked) species at a certain time t, given certain initial conditions, or the steady state solution of the equation for $\dot{\textbf{n}}=0$ \cite{HOLM2021100982, Greenland_Molecules}.

\subsection{Rate data}
\label{sec:rate_calc}
Before CRUMPET can be used, the relevant rate data and Einstein coefficients from equation \eqref{eq:Rate_eq} must be prepared to calculate the matrix $\mathbf{M}$. The reaction rates are given by 

\begin{equation}
    R^k_{i,j} = \langle\sigma_{i,j}^k(E)v\rangle,
\end{equation}

where $\sigma_{i,j}^k(E)$ is the cross section related to the collision of species $i$ and $j$ that results in species $k$, and $v$ is the collision velocity of the two species $i$ and $j$. 

Because much of the collisional data used in the CRM is available only in terms of cross sectional data, the cross sections must be converted to rate coefficients. This can be done using an integral over the Boltzmann distribution $f(v)$, which represents the velocity distribution of the electrons for electron-impact interactions.

\begin{equation}
    f(v) = \left(\frac{m}{2\pi k_B T_e}\right)^{3/2}4\pi v^2 e^{\frac{mv^2}{2k_B T_e}}.
    \label{eq:Bolzmann}
\end{equation}

Because deuterium molecules are over 7000 times more massive than electrons, their velocity will be negligible compared to that of the electrons. It is therefore sufficient to assume the case of a stationary molecule, hit by an electron at finite velocity. Therefore, for some cross section $\sigma$ the reaction rate can be calculated using 

\begin{equation}
    \langle\sigma v\rangle = \int^{\infty}_{0} f(v)v\sigma(v)dv.
    \label{eq:Bolzmann_integral}
\end{equation}

with $v=\sqrt{\frac{2E}{m_e}}$. Note that \eqref{eq:Bolzmann_integral} is not valid for reactions involving species with similar masses, for example for reactions between $D^+$ and $D_2$. In that case, a double integral must be performed over the velocity spaces of both species.  

The cross sectional data used in this report is primarily obtained from the Laporta \cite{Laporta_2021} and MCCCDB databases \cite{MCCCDB_exc_exc,MCCCDB_exc_X1,MCCCDB_Ionization}. These databases contain cross sections for particle collisions obtained from ab initio quantum mechanical calculations. The Einstein coefficients for spontaneous emission are used from Fantz et al. \cite{Fantz2002}. 

\subsection{Detailed balance}
\label{sec:det_balance}
The method of \autoref{sec:rate_calc} can be used to calculate the rate data as input for CRUMPET, from the available cross sectional data. However, for many reactions, only cross sections for excitation to higher energy states are available. For example for vibrational excitation in the ground state
\begin{equation}
    e+\text{D}_2(\nu=i)\rightarrow e+\text{D}_2(\nu=j)
    \label{eq:rea_vibr_trans}
\end{equation}
cross sections are only given for $j > i$ \cite{Laporta_2021}. The cross sections for deexcitation can be obtained using the principle of detailed balance, which states that in thermal equilibrium, the excitation and deexcitation rates must be equal. The collisional deexcitation cross sections can then be calculated as
\begin{equation}
    \sigma_{deex}(T) = \sigma_{ex}(T)e^{-\frac{\Delta E}{k_BT}},
    \label{eq:det_balance}
\end{equation}
where $\sigma_{deex}(T)$ and $\sigma_{ex}(T)$ are deexcitation and excitation cross sections respectively, and $\Delta E$ is the energy difference between the two states. Using the cross section from \eqref{eq:det_balance}, and equation \eqref{eq:Bolzmann_integral}, the collisional deexcitation rates can be obtained \cite{Greenland2001}. The principle of detailed balance can also be applied to other reactions, such as excitation and deexcitation between electronically excitated states, or vibrational transitions in excited states. 

When the principle of detailed balance is applied, and other reactions are switched off, the solution to equation \eqref{eq:matrices} is given by a Boltzmann of the form 
\begin{equation}
    n_k = Cg_k e^{-\frac{E_k}{k_BT}},
    \label{eq:bolzmann}
\end{equation}
where $n_k$ is the density of particles in state $k$, $C$ is a constant, $g_k$ is the degeneracy of state $k$, and $E_k$ is its energy. Note that the vibrationally excited states of molecular deuterium are non-degenerate, so $g_k=1$. 

\subsection{Effective rates}
Once the required input is generated, CRUMPET can be used to solve the system of equations \eqref{eq:matrices} to obtain the densities $\mathbf{n}$. This allows us to calculate the effective rates for certain reactions. To illustrate the calculation of effective rates, molecular charge exchange is used here as an example. In this example, we will consider only vibrationally excited molecules. 

\begin{equation}
    \text{D}^++\text{D}_2(\nu) \xrightarrow{\langle\sigma v\rangle_\nu} \text{D}+ \text{D}_2^+
    \label{eq:mol_cx}
\end{equation}

The molecular deuterium is in a certain vibrational state $\nu$. As stated in \autoref{sec:E_levels}, it is important to track these states, because the reaction rates $\langle\sigma v\rangle_\nu$ depend on the vibrational state $\nu$. The effective rate of molecular charge exchange is given by a weighted sum over the different $\langle\sigma v\rangle_\nu$:

\begin{equation}
    \langle\sigma v\rangle_{eff} = \sum^{\nu_{max}}_{\nu=0} f_\nu \langle\sigma v\rangle_\nu
    \label{eq:eff_rate}
\end{equation}

Here, $f_\nu = \frac{n_\nu}{n_{0}}$ is the fraction of all molecules that are in state $\nu$, $\langle\sigma v\rangle_\nu$ is the reaction rate of state $\nu$, $\nu_{max}$ is the maximum vibrational quantum number, and $\langle\sigma v\rangle_{eff}$ is the effective reaction rate. Since $f_\nu$ will in general depend on both electron density and temperature, so will $\langle\sigma v\rangle_{eff}$. The absolute reaction rate for molecular charge exchange (in m$^{-3}$s$^{-1}$) is then equal to $n_en_0\langle\sigma v\rangle_{eff}$, where $n_e$ is the electron density and $n_0$ is the total molecular density. The effective rates are important because these are widely used in larger 2D or 3D models. These computationally expensive codes often cannot afford to track each individual vibrationally excited state of all the molecules simultaneously, so they often use effective rate coefficients for the reactions.

\subsection{Modelling $D_2^*$ radiative emission}
CRUMPET can also be used to investigate radiative emission from the molecules. As stated in \autoref{sec:E_levels}, molecules can be excited to higher electronic states, which decay radiatively, leading to photon emission. It is therefore possible to use spectroscopy to analyse the rovibronic states of the molecules \cite{Fantz2002}.

We can analyse radiative emisison by using the output densities of the CRM (using \eqref{eq:matrices}) to calculate the line emission intensity of the transition from state $p$ to $q$ (in units of photons m$^{-3}$s$^{-1}$).

\begin{equation}
    I_{pq} = n_pA_{pq} = n_en_0\kappa_{p}A_{pq},
    \label{eq:line_intensity}
\end{equation}
where $n_p$ is the density of state $p$, $A_{pq}$ is the Einstein coefficient for spontaneous emission corresponding to the transition from $p$ to $q$, $n_e$ is the electron density and $n_0$ is the total density of deuterium molecules. $\kappa_p$ is the so-called population coefficient for state $p$, given by 
\begin{equation}
    \kappa_p = \frac{n_p}{n_en_0}.
    \label{eq:kappa_p}
\end{equation}

Effective line emission rate coefficients $X^{eff}_{pq}$ can be calculated, which can be tabulated as function of $n_e$, $T_e$ for ease of use in modelling molecular emission and analysing experimental spectra.
\begin{equation}
    X^{eff}_{pq} = \kappa_{p}A_{pq}.
    \label{eq:Xeff}
\end{equation}

The above equations were adapted from the Yacora H$_2$ model help page \cite{Yacora_on_the_Web}. We can also calculate the radiative power loss rate coefficient for each transition by multiplying $X^{eff}_{pq}$ with the energy corresponding to the transition. 
\begin{equation}
    P^{eff}_{pq} = \epsilon_{pq}X^{eff}_{pq},
\end{equation}
where $\epsilon_{pq}$ is the energy corresponding to the transition from $p$ to $q$. $P^{eff}_{pq}$ has dimensions of eVm$^3$s$^{-1}$. We can calculate the total radiative power losses in the model by summing over all the radiative transitions.

\subsection{Modelling vibrational distribution in upper Fulcher state.}
\label{sec:Fulcher_background}
The previous section shows how to use the densities obtained from the CRM to calculate line emission rate coefficients for the different transitions between electronic states. One transition that is of particular interest to spectroscopy is the transition from the $d^3\Pi_u$ to $a^3\Sigma_g$ state (d to a in \autoref{fig:E_diagram}). Light from this transition, also known as the Fulcher band, is in the visible range, which makes it easy to analyse using spectroscopic techniques. The molecules in the $d^3\Pi_u$ upper Fulcher state have their own vibrational and rotational distribution, and information about this distribution is encoded in the emitted Fulcher band. 

In literature, the vibrational distribution in the upper Fulcher state is often analysed by assuming a simple model for the vibrational distribution in the electronic ground state, which is then mapped to the $d^3\Pi_u$ state using Franck-Condon factors. 

Franck-Condon factors determine the probability of the radiative transition from a molecule D$_2(n=A,\nu=i)$ to D$_2(n=B,\nu=j)$. These factors can be used to estimate the vibrational distribution in the upper Fulcher state using the following formula.
\begin{equation}
    f_j^{d} = \sum_if^{X}_iK_{ij}.
    \label{eq:mapping}
\end{equation}
$f_j^{d}$ is the fraction of D$_2(d^3\Pi)$ molecules in vibrational state $j$, $f^{X}_i$ is the fraction of ground state molecules in vibrational state $i$, and $K_{ij}$ is the Franck-Condon factor corresponding to the radiative transition between D$_2(X^1\Sigma_g,\nu=i)$ and D$_2(d^3\Pi_u,\nu=i)$.

The advantage of this method is that there is no need to model the upper state, you only need the ground state distribution to calculate the populations in the upper state. This does not necessarily require a CRM: assuming the ground state vibrational distribution is populated according to a Boltzmann relation with a certain vibrational temperature, the vibrational distribution in the Fulcher state can be modelled and a vibrational temperature (if the measurement is in agreement with this model) can be inferred \cite{Fantz2002}. A possible limitation of this technique is the fact that the Franck-Condon factors only apply to radiative transitions, while excitation from ground to the upper Fulcher state is collisional. This mapping may therefore not be sufficiently accurate. Additionally, the vibrational distribution of the electronic ground state may not always be described by a Boltzmann relationship. The CRM of this report also allows us to calculate the vibrational distribution in the upper Fulcher state directly, in stead of using the mapping of equation \eqref{eq:mapping}.

\subsection{Investigation of Eirene effective and vibrationally resolved rates.}
\label{sec:Eirene_rates}
\begin{figure}[h]
    \centering
    \includegraphics[width=0.5\textwidth]{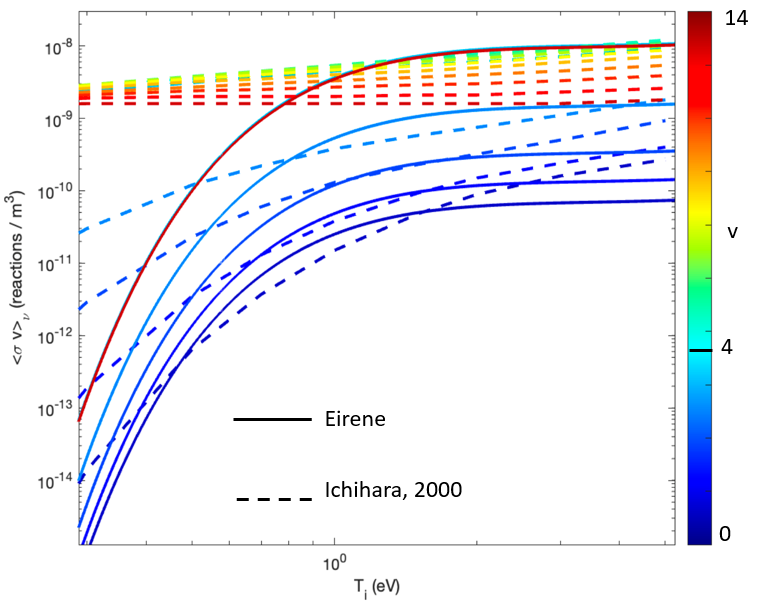}
    \caption{Comparison between the vibrationally resolved rate contstants for molecular charge exchange from Eirene \cite{Eirene} and Ichihara et al. \cite{Ichihara_2000}. The different lines represent rate coefficients for the different vibrational levels of hydrogen. Credit for this image goes to  Dr. Kevin Verhaegh. }
    \label{fig:vibr_res_CX}
\end{figure}

Before diving into the results of this report, the default tabulated Eirene rate data and its inaccuracies are investigated. As stated in the introduction, there are large concerns with regards to the accuracy of Eirene rate data for molecules, commonly used as input data for SOLPS simulations \cite{Verhaegh_2023_SOLPS}. In 2021, Verhaegh et al. used spectroscopic techniques to investigate TCV divertor plasmas, and obtained results that deviate from the SOLPS simulations \cite{Verhaegh_2021}. Preliminary results show similar discrepancies between experiment and simulations are found in the MAST-U Super-X divertor \cite{verhaegh2023role}. It is expected that much of this difference is due to the underestimation of the molecular charge exchange (equation \eqref{eq:mol_cx}) rates by Eirene \cite{Verhaegh_2023_SOLPS,Verhaegh_2021,verhaegh2023role}. A comparison of vibrationally resolved charge exchange rates of molecular hydrogen between the tabulated vibrationally resolved Eirene rates ('H2VIBR'), and the rates calculated by Ichihara et al. in 2000 is given in \autoref{fig:vibr_res_CX} \cite{Eirene,Ichihara_2000}.
The rates from the Eirene documentation are vastly different from the Ichihara rates. This is partly due to the method that was used for the calculation of rate coefficients at $\nu>0$ in Eirene.  The molecular charge exchange rates for $\nu=0$ in 'H2VIBR' were obtained experimentally from Holliday, et al. 1971 \cite{Holliday1971,Janev1987}. The rates at vibrationally excited levels are simply assumed to be equal to the ground state rate rate, multiplied by a scalar using a resonance scaling obtained from \cite{Greenland2001}. This assumption becomes problematic at lower temperatures because the threshold energy for the reaction is different for different vibrational levels, which is not accounted for with simple scalar multiplication. You can see the problem with this assumption in \autoref{fig:vibr_res_CX}. The Ichihara rates for $\nu\leq4$ at lower temperatures are almost independent of temperature, whereas the Eirene rates have the same temperature dependence as the ground state, only shifted up by a certain factor. This is expected to have a large impact on the effective charge exchange rates at lower temperatures. 

\begin{figure}[h]
     \centering
     \begin{subfigure}[b]{0.49\textwidth}
         \centering
         \includegraphics[width=\textwidth]{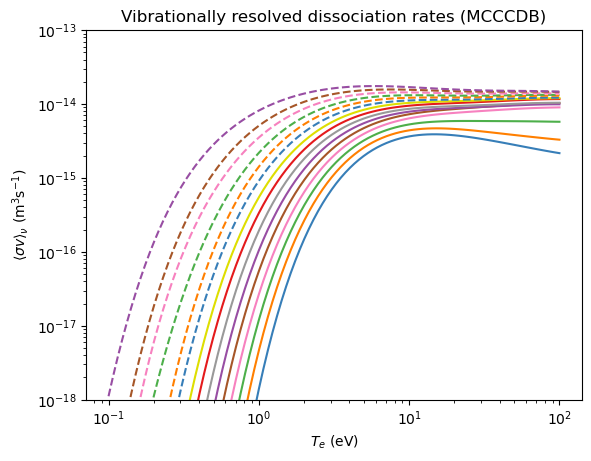}
         \caption{}
         \label{fig:Diss_MCCCDB}
     \end{subfigure}
     \hfill
     \begin{subfigure}[b]{0.49\textwidth}
         \centering
         \includegraphics[width=\textwidth]{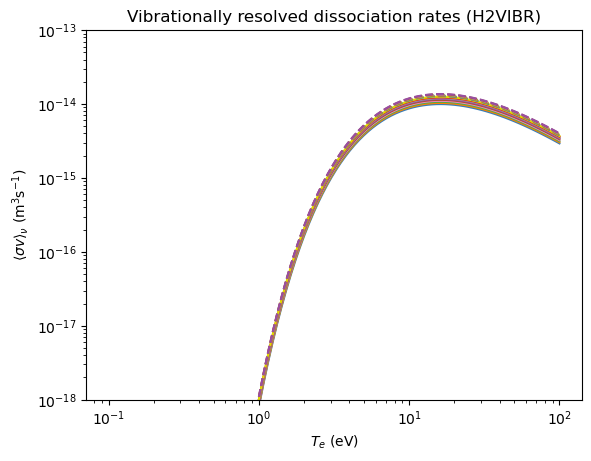}
         \caption{Molecular activated reaction rates via D$_2^+$.}
         \label{fig:Diss_Eirene}
     \end{subfigure}
        \caption{Vibrationally resolved rate coefficients for electron impact dissociation of hydrogen. (a) shows the rates calculated using MCCCDB cross sections and equation\eqref{eq:Bolzmann_integral} \cite{MCCCDB_exc_X1}. (b) contains the rates from the Eirene documentation \cite{Eirene}. }
        \label{fig:Diss_comparison}
\end{figure}

The method of simply multiplying the ground state rate coefficients with a scalar to obtain the rates for higher vibrational levels is used also for other reactions in the Eirene documentation, such as electron impact dissociation (equation \ref{eq:diss}), electron-impact ionisation and vibrational excitation through electron-impact. 

\begin{equation}
    e+\text{D}_2(\nu)\rightarrow e+2\text{D}
    \label{eq:diss}
\end{equation}
Rates for electron-impact dissociation, calculated using cross sections from the Molecular Close Convergent Coupling Database (MCCCDB), are compared to rates from H2VIBR in \autoref{fig:Diss_comparison} \cite{MCCCDB_exc_X1} \cite{MCCCDB_exc_X1}. The tabulated 'H2VIBR' rate for $\nu=0$ was obtained from \cite{Janev1987,Takayanagi1977}. The 'H2VIBR' dissociation rates assume dissociation occurs after electronic excitation to the $b3\Sigma_u^+$, $a3\Sigma_g^+$ and $c3\Pi_u$ electronic states. For the H2VIBR rates, there is no large difference between different vibrational levels, whereas the MCCCDB data shows large variations for different $\nu$. Again the Eirene data is obtained by scalar multiplication from the vibrational ground state \cite{Greenland2001}, while the MCCCDB cross sections are calculated using sophisticated quantum mechanical methods. The result is that dissociation using H2VIBR rates is virtually impossible below 1 eV, which is not the case for the higher vibrational levels from MCCCDB.

Eirene generally employs two different databases with tabulated polynomial fits for plasma-molecular interactions. The 'H2VIBR' tables provide vibrationally resolved rates, which was employed in Eirene to model the vibrational distribution. It is also used in SOLPS-ITER simulations that track every vibrational level individually \cite{Wischmeier2005}. Eirene also provides polynomial fits for (vibrationally unresolved) effective rate coefficients in 'AMJUEL', which are the rates typically used in plasma-edge simulations. The 'AMJUEL' and 'H2VIBR' rates for electron-impact dissociation and molecular ionisation are not self-consistent. Vibrationally resolved electron-impact dissociation and molecular ionisation rates ('H2VIBR') were obtained for electron-impact collisions with the electronic ground state \cite{Janev1987,Takayanagi1977}. In contrast, the 'AMJUEL' rates were obtained from an electronically resolved (but vibrationally unresolved) CRM for the vibrational ground state \cite{Sawada1995}. The same analytic re-scaling as employed in 'H2VIBR' \cite{Greenland2001}, combined with a vibrational distribution modelled using the rates in 'H2VIBR', was then used to compute effective rates. 

The effective rates in AMJUEL are calculated for hydrogen. To adjust for the difference in isotope mass for deuterium plasmas, the common strategy in SOLPS simulations is to re-scale the temperature with the isotope mass for reactions that involve collisions with ions \cite{Verhaegh_2023_SOLPS}. 
\begin{equation}
    \langle\sigma v\rangle_{eff,\text{D,Eirene}}(T) = \langle\sigma v\rangle_{eff,\text{H,Eirene}}(T/2).
    \label{eq:Eirene_rescaling}
\end{equation}
The effective rate coefficients for ion collisions for deuterium at temperature $T$ is assumed to be equal to the effective rates for hydrogen at $T/2$. The problem with this re-scaling is that according to equation \eqref{eq:eff_rate}, the effective rate also depends on the vibrational distribution of the molecules, which is a function of temperature. Therefore, by applying equation \eqref{eq:Eirene_rescaling} to the effective rates, the vibrational distribution $f_\nu(T)$ is also rescaled. The following equations show the Eirene and correct rates for deuterium respectively \cite{Verhaegh_2023_SOLPS}. 
\begin{align}
    \langle\sigma v\rangle_{eff,\text{D,Eirene}}(T) = \sum_{\nu=0}^Nf_\nu(T/2)\langle\sigma v\rangle_{\nu}(T/2) \\
    \langle\sigma v\rangle_{eff,\text{D,Correct}}(T) = \sum_{\nu=0}^Nf_\nu(T)\langle\sigma v\rangle_{\nu}(T/2)
    \label{Eirene_rescaling}
\end{align}

In conclusion, the combined issues of isotope mass rescaling of effective rates and inaccurate assumptions about the vibrationally resolved rates result in incorrect estimations of the importance of molecular processes in deuterium plasmas. The decrease of threshold energies for higher vibrational levels leads to higher rate constants at low temperatures for both dissociation and molecular charge exchange. As is shown in \autoref{sec:rates}, the correction of these errors leads to a large difference in the effective rates for dissociation, and especially for molecular charge exchange.

%% file: 3_Results.tex
\section{Results}
\label{sec:Results}

\subsection{The model setup}
\label{sec:model}
This section contains a summary of the setup of the model used in this report. As previously mentioned, the CRM contains regular (tracked) and background species. The background species are electrons, D$^+$ ions and D atoms. The densities of these species are kept constant. The other species include  D$_2^+$ and D$^-$ ions, as well as vibronically (i.e. vibrationally and electronically) excited D$_2$ molecules. The vibronic ground state of diatomic deuterium D$_2(X^1\Sigma_g, v=0)$ is also included as a background species, with constant density. Each of the excited states of D$_2$ is vibrationally resolved up to and including vibrational quantum number $\nu=14$. As stated in \autoref{sec:E_levels}, rotational levels are not included in this model. 

The excited states included in the model are the $X^1\Sigma_g$, $B^1\Sigma_u$, $C^1\Pi_u$, $EF^1\Sigma_g$, $a^3\Sigma_g$, $c^3\Pi_u$ and $d^3\Pi_u$ states. In \autoref{fig:E_diagram}, these states correspond to the energy levels marked X, B, C, EF, a, c and d. The $b^3\Sigma_u$ state (b in \autoref{fig:E_diagram}) is not explicitly included in the model because it is a repulsive state. Any molecule that ends up in this state therefore immediately dissociates. The $d^3\Pi_u$ state is included because the transition from $d^3\Pi_u$ to $a^3\Sigma_g$ causes the visible Fulcher emission (see \autoref{sec:Fulcher_background}).

\autoref{tab:rea_ref} contains a summary of the different reactions included in the model, as well as the databases used to obtain the necessary data. For a complete list with all the different transitions, as well as the corresponding references, see \autoref{sec:reaction_list}. \autoref{sec:rate_explanation} has a detailed explanation of how the rates for the reactions of \autoref{tab:rea_ref} are calculated.

\begin{table}[H]
\centering
\begin{tabular}{|l|l|l|}
\hline
Reaction                & Formula & Database \\
\hline
 & & \\
Vibrational transitions & $e+\text{D}_2(n,\nu=i)\rightarrow e+\text{D}_2(n,\nu=j)$    &  Laporta  \cite{Laporta_2021}       \\
 & & \\
Dissociative attachment & $e+\text{D}_2(n,\nu)\rightarrow \text{D}^-+\text{D}$       &  Laporta  \cite{Laporta_2021}    \\
 & & \\
Ionization & $e+\text{D}_2(n,\nu)\rightarrow 2e+\text{D}_2^+$ & MCCCDB \cite{MCCCDB_Ionization}\\
 & & \\
Electronic excitation & $e+\text{D}_2(n=A,\nu=i)\rightarrow e+\text{D}_2(n=B,\nu=j)$ & MCCCDB \cite{MCCCDB_exc_X1,MCCCDB_exc_exc}\\
 & & \\
Radiative deexcitation & $\text{D}_2(n=A,\nu=i)\rightarrow \text{D}_2(n=B,\nu=j) + \hbar\omega$ & Fantz \cite{FANTZ2006853}\\
 & & \\
Electron impact dissociation & $e+\text{D}_2(n,\nu)\rightarrow e+2\text{D}$ & MCCCDB \cite{MCCCDB_exc_X1}\\
 & & \\
Molecular charge exchange & $\text{D}^++\text{D}_2(n,\nu)\rightarrow \text{D} + \text{D}_2^+$ & Ichihara \cite{Ichihara_2000}\\
 & & \\
Dissociative recombination & $e+\text{D}_2^+\rightarrow 2\text{D}$ & AMJUEL \cite{Eirene} \\
 & & \\
Electron impact dissociation (with D$_2^+$) &  $e+\text{D}_2^+\rightarrow e + \text{D}^+ + \text{D}$ & AMJUEL \cite{Eirene}\\
 & & \\
Dissociative ionization (with D$_2^+$) &  $e+\text{D}_2^+\rightarrow 2e + 2\text{D}^+$ & AMJUEL \cite{Eirene} \\
 & & \\
Charge exchange &  $\text{D}^++\text{D}^-\rightarrow 2\text{D}$ & AMJUEL \cite{Eirene} \\
 & & \\
Ionization (with D$^-$) &  $\text{D}^++\text{D}^-\rightarrow 2e + \text{D}^+ + \text{D}$ & AMJUEL \cite{Eirene} \\
 & & \\
\hline
\end{tabular}
\caption{References for reaction data}
\label{tab:rea_ref}
\end{table}

What is novel about this particular model, is that all the electronically excited molecules are vibrationally resolved. This is not the case for most models. For example, Yacora-on-the-Web does not include vibrational levels in its CRM \cite{Yacora_on_the_Web}. Another example is that electronic excited levels are not included in the vibrationally resolved Eirene rates used for modelling the vibrational distribution to compute effective rates; whereas coupling between vibrational and electronic excited levels was not properly included for calculating the effective Eirene rates. Since many reaction rates depend greatly on the vibrational states of the molecules, it is important to monitor the abundance of molecules in different vibrational states. Spontaneous radiative decay from electronic excited levels can redistribute the vibrational distribution in the electronic ground state. The current model allows the calculation of the vibrational distribution, not only in the electronic ground state, but the excited states as well. 

Additionally, the collisional and radiative data used to generate the CRM is mostly comprised of accurate quantum-mechanical calculations of cross sections and transition probabilities that are fully vibrationally resolved: they do not rely on analytically rescaling rates from the vibrational ground state to higher levels. Most of this data is given also for deuterium, in stead of just hydrogen. Previously used data, such as in the Eirene documentation, is only given for H. When doing calculations with deuterium, simple ion temperature mass scaling is applied, which was shown to be inaccurate. Thus, by incorporating cross sections and radiative transition probabilities calculated specifically for deuterium, it is possible to calculate more accurate distributions and effective rates. 

\subsection{Effective reaction rates compared to AMJUEL rates.}
\label{sec:rates}
\begin{figure}[h]
    \centering
    \includegraphics[width=0.5\textwidth]{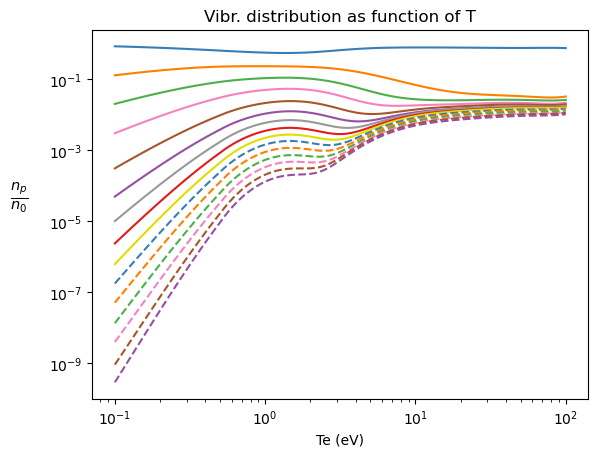}
    \caption{Vibrational distribution of molecules in the ground state as a function of temperature. The different lines represent the different vibrational states. The data in this figure was calculated using $n_e=10^{19}$ m$^{-3}$.}
    \label{fig:vibr_dist_full}
\end{figure}

As stated in \autoref{sec:Eirene_rates}, the reaction rate coefficients currently used in SOLPS are somewhat inaccurate. In this section, the model described in \autoref{sec:model} is used to generate alternative rate coefficients, which are then compared to the default AMJUEL rates (\cite{Eirene}), and to experimental results in strongly detached divertor plasmas in MAST-U. The vibrational distribution as a function of temperature is given in \autoref{fig:vibr_dist_full}. At low temperatures, the distribution roughly follows a Bolzmann distribution, and there are very few vibrationally excited molecules. At higher temperatures, the number of vibrationally excited molecules increases. 

\begin{figure}[h]
     \centering
     \begin{subfigure}[b]{0.49\textwidth}
         \centering
         \includegraphics[width=\textwidth]{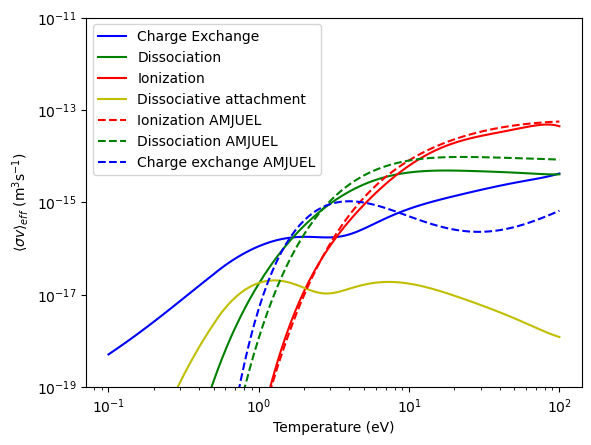}
         \caption{Reaction rates}
         \label{fig:rates_n=19}
     \end{subfigure}
     \hfill
     \begin{subfigure}[b]{0.49\textwidth}
         \centering
         \includegraphics[width=\textwidth]{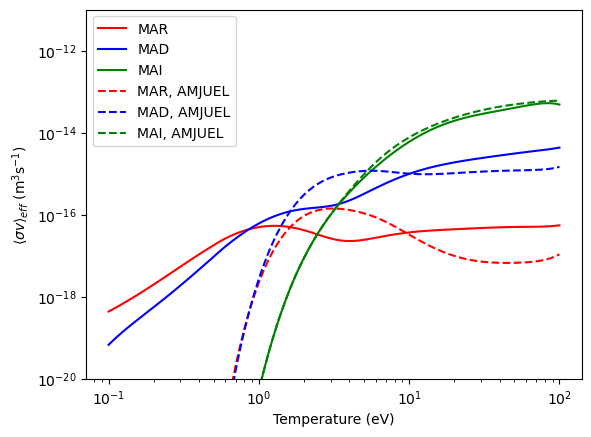}
         \caption{Molecular activated reaction rates via D$_2^+$.}
         \label{fig:MA_plus_n=19}
     \end{subfigure}
        \caption{(a) contains reaction rates for molecular charge exchange, dissociation, ionization and dissociative attachment, compared to rates from AMJUEL \cite{Eirene}. (b) contains the rate coefficients for MAR, MAI and MAD via D$_2^+$ ions, also compared to tabulated rates from AMJUEL \cite{Eirene}. The rates in this figure were calculated for an electron density $n_e=10^{19}$ m$^{-3}$}
        \label{fig:rates_MA_plus}
\end{figure}

\autoref{fig:rates_n=19} contains a plot of the reaction rate coefficients for charge exchange, dissociation, ionization and dissociative attachment. The rates are compared to rates determined from the tabulated polynomial fit coefficients in AMJUEL. The rates for ionization are quite close to the tabulated data, but the rates for dissociation and especially charge exchange differ greatly from the AMJUEL data, especially at low temperatures. 

The effective rates calculated using the model of \autoref{sec:model} were also compared to rates calculated using vibrationally resolved rates from the Eirene documentation H2VIBR. The discrepancy between the effective D$_2$ dissociation rates from our calculations and those obtained from H2VIBR are greater than those obtained from AMJUEL. This shows that these rates in H2VIBR and AMJUEL are not self-consistent. This implies that previous vibrationally resolved plasma-edge simulations (which use H2VIBR) are inconsistent with simulations that use effective rates (which use AMJUEL), in part because the rates themselves are inconsistent.

The reactions of \autoref{fig:rates_n=19} are the first step in co-called molecular activated reactions. These reactions are multistep processes that involve molecules, and can form ion sources/sinks and neutral atom sources. In the Molecular Activated Recombination (MAR) reaction chain, for example, a plasma ion gets neutralised. MAR is, therefore, an ion sink. Molecular Activated Dissociation (MAD) reaction chains provide additional mechanisms for dissociating molecules into neutral atoms beyond direct electron-impact dissociation and is thus a neutral atom source. Molecular Activated Ionization (MAI) reaction chains result in the ionisation of neutral atoms and thus acts as an ion source. MAR and MAD can occur both via the creation of D$_2^+$ and $D_2^- \rightarrow$ D$^- + D$ ions. 

As demonstrated by \autoref{fig:rates_n=19}, the molecular charge exchange rate at low temperatures is highly underestimated in the AMJUEL rates. This leads to large underestimation of the amount of D$_2^+$ in the plasma, and therefore the importance of MAR and MAD at low temperatures. \autoref{fig:MA_plus_n=19} contains a plot of the MAR, MAD and MAI rates via D$_2^+$, compared to rates calculated using data from AMJUEL. The MAR and MAD rates at low temperatures are much higher than the AMJUEL rate data suggests. Molecular processes, therefore, can play a much larger role in the plasma at low temperatures than expected.

\begin{figure}[H]
     \centering
     \begin{subfigure}[b]{0.49\textwidth}
         \centering
         \includegraphics[width=\textwidth]{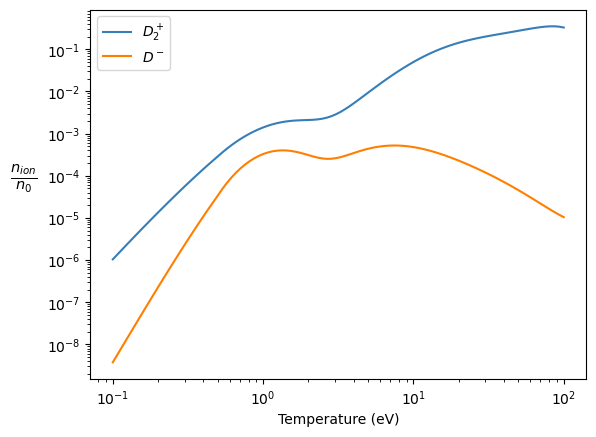}
         \caption{Ion density fractions}
         \label{fig:ions}
     \end{subfigure}
     \hfill
     \begin{subfigure}[b]{0.49\textwidth}
         \centering
         \includegraphics[width=\textwidth]{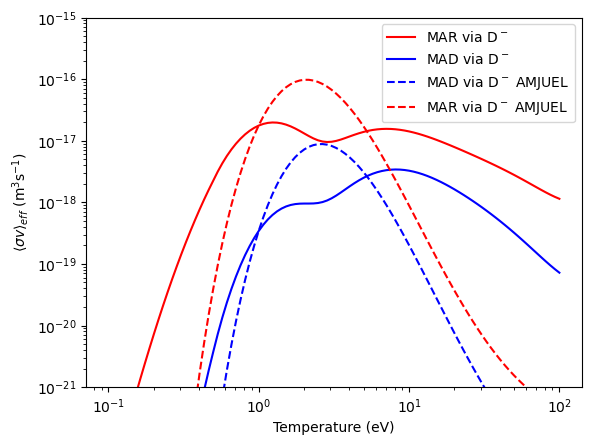}
         \caption{Molecular activated reaction rates via D$^-$.}
         \label{fig:MA_min_n=19}
     \end{subfigure}
        \caption{(a) contains the ion densities divided by the total molecular density as a function of temperature. (b) contains the rate coefficients for MAR and MAD via D$^-$ ions, compared to rates obtained from the tabulated polynomial fit coefficients from AMJUEL \cite{Eirene}. The data in this figure was calculated for an electron density $n_e=10^{19}$ m$^{-3}$.}
        \label{fig:rates_MA_min}
\end{figure}

MAR and MAD may also occur via $D^-$. Although such rates are available in AMJUEL for hydrogen, the applicability of these rates for deuterium has been extensively debated \cite{Kukushkin2017,Verhaegh_2021,Reiter2018,Krishnakumar2011}, since the cross-section for dissociative attachment has been measured to be orders of magnitude lower for deuterium than for hydrogen for the vibrational ground state \cite{Krishnakumar2011}. It is, however, the higher vibrational levels that drive dissociative attachment, where the isotope difference is reduced. \autoref{fig:MA_min_n=19} demonstrates that, although the calculated MAR and MAD rates via D$^-$ differ greatly from the AMJUEL $H^-$ rates, MAR through $D^-$ may be non-negligible.  

Using the model, it is also possible to calculate the amount of D$_2^+$ and D$^-$ in the plasma assuming no transport of these ions occurs. \autoref{fig:ions} contains the densities of the different ions in the plasma divided by the total molecular density as a function of temperature. These ratios are useful for calculating MAR, MAD and MAI rates as well as for modeling the hydrogenic emission arising from excited atoms after molecular break-up \cite{Verhaegh_2021}.

\begin{figure}[h]
     \centering
     \begin{subfigure}[b]{0.49\textwidth}
         \centering
         \includegraphics[width=\textwidth]{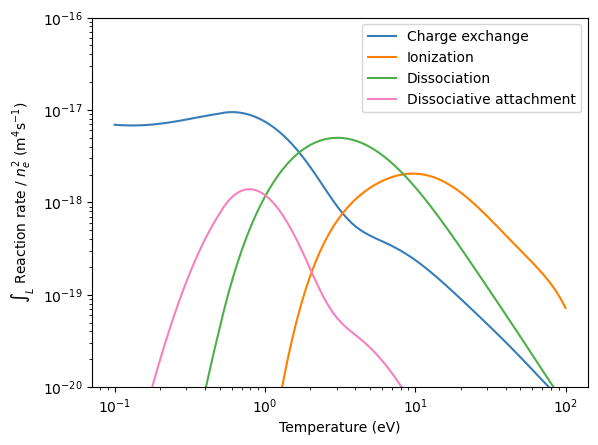}
         \caption{}
         \label{fig:reactions_int}
     \end{subfigure}
     \hfill
     \begin{subfigure}[b]{0.49\textwidth}
         \centering
         \includegraphics[width=\textwidth]{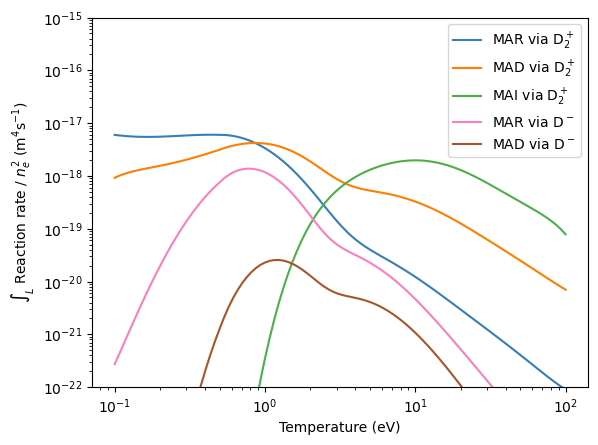}
         \caption{}
         \label{fig:MA_reactions_int}
     \end{subfigure}
        \caption{Plots of the reaction rate (in m$^{-3}$s$^{-1}$) integrated over a line of sight $L$, divided by $n_e^2$. (a) contains the rates for charge exchange, dissociation, ionization and dissociative attachment. (b) contains the MAR, MAD and MAI rates for both D$^+_2$ and D$^-$ ions. The data in this figure was calculated using the scaling law for $f(T)$ for MAST-U, and using an electron density $n_e=10^{19}$ m$^{-3}$.}
        \label{fig:rates_int}
\end{figure}

The impact of the high effective reaction rate for molecular charge exchange and dissociation at low temperatures is even more pronounced when the molecular density as a function of temperature is considered. Since molecular density increases at low temperatures \cite{Stangeby2017,Verhaegh_2023_fdens}, the impact of plasma-molecular interactions will be even larger. We can use a toy model to visualise this by using a scaling law $f_{\text{D}_2}(T)$, generated by Verhaegh et al. for the MAST-U Super-X divertor \cite{Verhaegh_2023_fdens} to approximate the molecular density as function of electron temperature. This scaling law was determined for MAST-U using synthetic spectroscopic techniques applied to SOLPS-ITER simulations, and can be used to estimate the reaction rate integrated along the line of sight $L$. 

\begin{equation}
    \int_LRdl = n_e\langle\sigma v\rangle_{eff}\int_L n_0dl = f_{\text{D}_2}(T)n_e^2\langle\sigma v\rangle_{eff} 
\end{equation}
where $R$ is the reaction rate (in m$^{-3}$s$^{-1}$), and $f_{\text{D}_2}(T)$ is a factor that relates the molecular density integrated over the line of sight to the electron density by 
\begin{equation}
    f_{\text{D}_2}(T) = \frac{1}{n_e}\int_L n_0(T,l)dl.
\end{equation}
We can use $f_{\text{D}_2}(T)$ to estimate the effect of molecular reactions, while taking $n_0$ as a function of temperature into account. The formula for $f_{\text{D}_2}(T)$ is given in \autoref{sec:scalings}. \autoref{fig:rates_int} depicts the rates integrated over a line of sight L, calculated using $f_{\text{D}_2}(T)$. \autoref{fig:reactions_int} shows a decrease of molecular reactions for high temperatures. For low temperatures however, the charge exchange rate seems to remain roughly constant. This leads to a large D$_2^+$ density, and to large MAR and MAD, which can be seen in \autoref{fig:MA_reactions_int}.

\begin{figure}[h]
     \centering
     \begin{subfigure}[b]{0.49\textwidth}
         \centering
         \includegraphics[width=\textwidth]{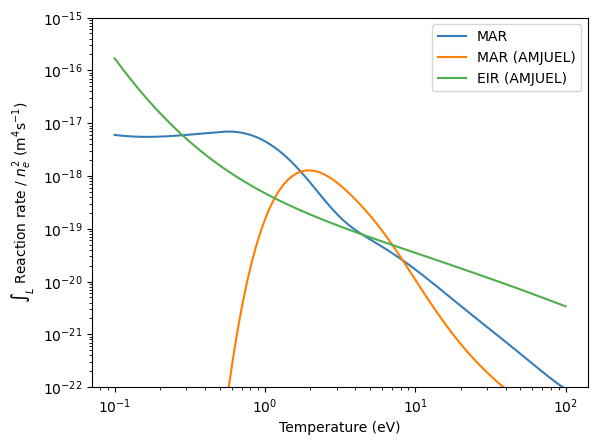}
         \caption{$n_e=10^{19}$ m$^{-3}$}
         \label{fig:ion_sinks_n=19}
     \end{subfigure}
     \hfill
     \begin{subfigure}[b]{0.49\textwidth}
         \centering
         \includegraphics[width=\textwidth]{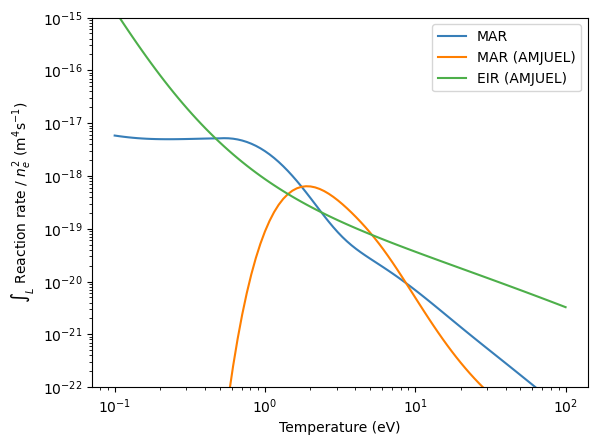}
         \caption{$n_e=10^{20}$ m$^{-3}$}
         \label{fig:ion_sinks_n=20}
     \end{subfigure}
        \caption{Comparison of different ion sinks MAR and EIR. Since atomic states are not explicitly tracked in the model of \autoref{sec:model}, the EIR data showed is calculated using standard AMJUEL data. Also MAR data from AMJUEL is included \cite{Eirene}. The comparison is shown for $n_e=10^{19}$ m$^{-3}$ and $10^{20}$ m$^{-3}$. The data in this figure was calculated using the scaling laws for D and $D_2$ from \cite{Verhaegh_2023_fdens}}
        \label{fig:ion_sinks}
\end{figure}

It is possible to use this toy model to investigate the relative importance of molecular processes compared to atomic ones by using scaling laws for the molecular and atomic densities. Recombination of electrons and ions can occur in two different ways: MAR, which involves molecules, and Electron Ion Recombination (EIR), which can be two and three-body.
\begin{align*}
        e+\text{D}^+ &\rightarrow \text{D} \\
    e+e+\text{D}^+ &\rightarrow e+\text{D} 
\end{align*}
By using different scaling laws for the D$^+$ and D$_2$ densities, the relative importance of MAR and EIR as ion sinks is investigated. \autoref{fig:ion_sinks} compares EIR rates from AMJUEL and MAR rates from the model of \autoref{sec:model}. MAR calculated using AMJUEL rates are also included. The ion sink MAR is insignificant at low temperatures compared to EIR when AMJUEL data is used. With the model of \autoref{sec:model}, however, MAR can indeed be significant at low temperatures. This illustrates that molecular processes can play a much larger than is expected from Eirene data, in agreement with experiments \cite{Verhaegh_2021,Verhaegh_2023_fdens}. At higher electron densities, EIR increases in magnitude and the MAR/EIR ratio is reduced, as is shown in \autoref{fig:ion_sinks_n=20}, although MAR could still remain non-negligible.

An important caveat is that these scaling laws have been specifically obtained from SOLPS-ITER simulations of the MAST-U divertor, where the impact of molecular interactions is relatively more important due to the tightly baffled Super-X divertor chamber \cite{Verhaegh_2023_fdens}. The impact of molecular interactions, particularly at higher densities, would likely be less important when scaling laws of a more open divertor geometry are used, for instance the unbaffled TCV ones from \cite{Verhaegh_2023_fdens}.

The calculated rates can also be compared to experiment. Verhaegh et al. demonstrated in 2023 a greatly increased impact of plasma-molecular interactions leading to increased MAR rates and a reduced target ion flux in highly detached plasmas in the MAST-U Super-X divertor \cite{verhaegh2023role}. The highly detached conditions Verhaegh's paper correspond to the low temperature (below 1 eV) region in Figures \ref{fig:rates_MA_plus}, \ref{fig:rates_MA_min}, \ref{fig:rates_int} and \ref{fig:ion_sinks}. The toy model therefore is in qualitative agreement with experimental data in that MAR plays a much larger role in highly detached plasmas than currently used rate data suggests. Even quantitative agreement exists: the toy model predicts line-integrated MAR ion sinks of around $4\times10^{21}$ part/s/m$^2$ (assuming $n_e = 2 \cdot 10^{19}$ m$^{-3}$), which agree with experimental inferences (figure 4 of \cite{verhaegh2023role}).

In conclusion, the CRM setup discussed in \autoref{sec:model} produces effective reaction rates for plasma molecular interactions, in particular MAR and MAD, that are much higher for low temperatures than the AMJUEL rates. \autoref{fig:MA_reactions_int} demonstrates that the MAR rate remains high even at temperatures well below 1 eV. This theoretical prediction agrees with experiment. Using a toy model, \autoref{fig:ion_sinks} shows that MAR can be a significant contributor to the total recombination in the plasma, even at low temperatures. However, MAR does become relatively less important for higher electron densities. When Eirene data files, such as AMJUEL, are used in plasma edge models, the importance of MAR and MAD through $D_2^+$ is therefore massively underestimated for detached plasmas. This could lead to incorrect estimations for the ion and heat flux to the target in present-day, and possibly future, devices. It is therefore essential to revise the available rate data to include more recent calculations with improved accuracy. 

\subsection{Investigation of electronically excited states.}
\label{sec:exc}
As stated in \autoref{sec:model}, the model of this report includes 7 vibrationally resolved electronic states. With 15 different vibrational states each, this comes down to 105 different species just for molecular deuterium. For a CRM, this is still manageable, but it might be too computationally expensive to include all these states in vibrationally resolved SOLPS setups. It is therefore interesting to see whether including all the excited states in the model is necessary, or whether approximations that reduce the number of species in the model suffice. 

\begin{figure}[h]
     \centering
     \begin{subfigure}[b]{0.49\textwidth}
         \centering
         \includegraphics[width=\textwidth]{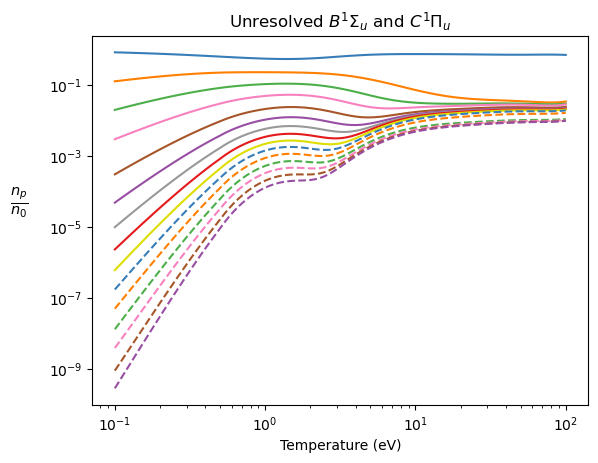}
         \caption{}
         \label{fig:vibr_dist_unr}
     \end{subfigure}
     \hfill
     \begin{subfigure}[b]{0.49\textwidth}
         \centering
         \includegraphics[width=\textwidth]{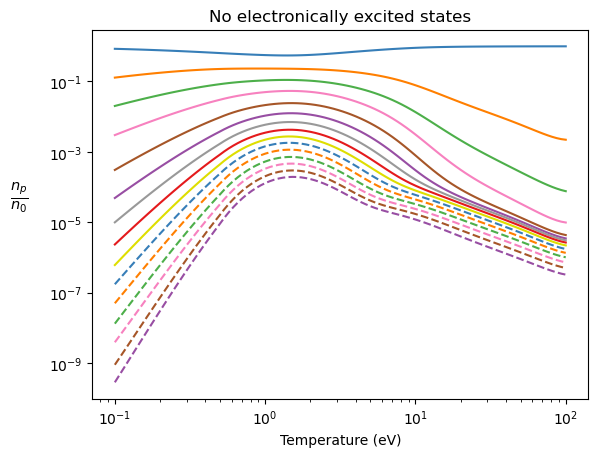}
         \caption{}
         \label{fig:vibr_dist_no}
     \end{subfigure}
        \caption{Vibrational distributions as a function of temperatures. (a) is for the case that only the $B^1\Sigma_u$ and $C^1\Pi_u$ states are included vibrationally unresolved. (b) contains the case without electronically excited states.}
        \label{fig:vibr_dist_unr+no}
\end{figure}

\autoref{fig:vibr_dist_unr+no} contains the vibrational distributions as a function of temperature, for the case with unresolved B and C states, and the case without electronically excited states. \autoref{fig:vibr_dist_unr} is quite similar to \autoref{fig:vibr_dist_full}, which indicates that vibrationally resolving the excited states has little impact on the vibrational distribution in the ground state. The distribution of \autoref{fig:vibr_dist_no} is also quite similar to the full model at low temperatures. For temperatures above 2 eV, the distribution without excited states is quite different than in \autoref{fig:vibr_dist_full}, presumably because at these temperatures electronic excitation becomes significant. 

To investigate the effect of electronically excited states on the molecular processes in the plasma, I will consider four different variations of the model of \autoref{sec:model}. The first one is the full model, with all 105 molecular states. The second is a model where only the electronic ground state, and the $B^1\Sigma_u$ and $C^1\Pi_u$ are included vibrationally resolved. This version thus has 3 electronic states, which gives 45 vibronic states. The third version is a model that has only has the ground state vibrationally resolved, and the $B^1\Sigma_u$ and $C^1\Pi_u$ states unresolved. This version therefore includes 15 vibrational states of the electronic ground state, and only 1 species for each of the excited states, giving 17 molecular species in total. The final model does not include excited states at all, so only 15 species are needed for the molecules.

\begin{figure}[h]
     \centering
     \includegraphics[width=0.5\textwidth]{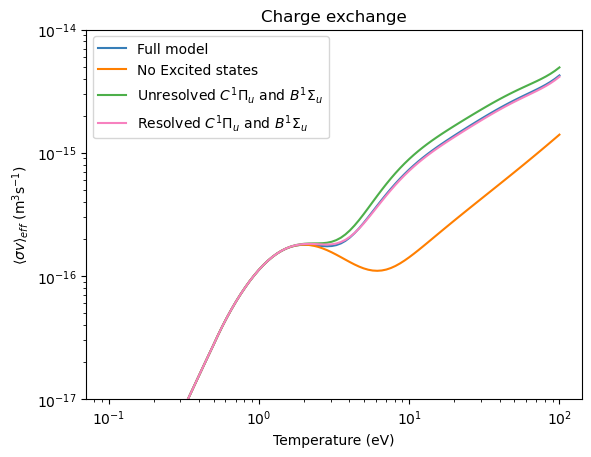}
     \caption{Effective molecular charge exchange rate as a function of temperature for the four different variations of the model.}
     \label{fig:cx_excited_states}
\end{figure}

A comparison between the different variations is given in \autoref{fig:cx_excited_states}. The charge exchange rates are all roughly the same below 2 eV. Above 2 eV, however, the case without electronic states becomes much lower than the others. This is because at these temperatures the redistribution of molecules in the ground state via excited states is significant. Without excited states, ionization and dissociation cause the depletion of vibrationally excited states, which leads to fewer vibrationally excited molecules (see \autoref{fig:vibr_dist_no}), and therefore lower charge exchange rates. The variation with resolved $B^1\Sigma_u$ and $C^1\Pi_u$ states gives a very similar result to the full model, which suggests that the states above $C^1\Pi_u$ do not affect the effective reaction rates significantly. The case with unresolved excited states is also quite similar to the full model, which suggests that vibrationally resolving the excited states might not be necessary. Similar comparisons to \autoref{fig:cx_excited_states} for dissociation, ionization, and dissociative attachment yield very similar results. 


The analysis of \autoref{fig:cx_excited_states} has shown that for the calculation of effective rates, it is not necessary to include all the excited molecular states vibrationally resolved. For temperatures below 2 eV, the excited states have no measurable effect. Above this temperature, excited states cause the redistribution of molecules in the ground state, thereby having an effect on the effective rates of the reactions. The excited state molecules themselves comprise less than 1\% of the total number of molecules in the plasma, so their impact on the effective rates is quite small. 

In conclusion, at higher temperatures, it is necessary to include electronic excitation in the model to accurately calculate the effective reaction rates. It is, however, not necessary to include states above $C^1\Pi_u$, and depending on the required accuracy, you might be able to get away with vibrationally unresolved electronically excited states.

\subsection{Emission spectra and radiative power loss.}
\label{sec:radiative_loss}
\begin{figure}[h]
     \centering
     \begin{subfigure}[b]{0.49\textwidth}
         \centering
         \includegraphics[width=\textwidth]{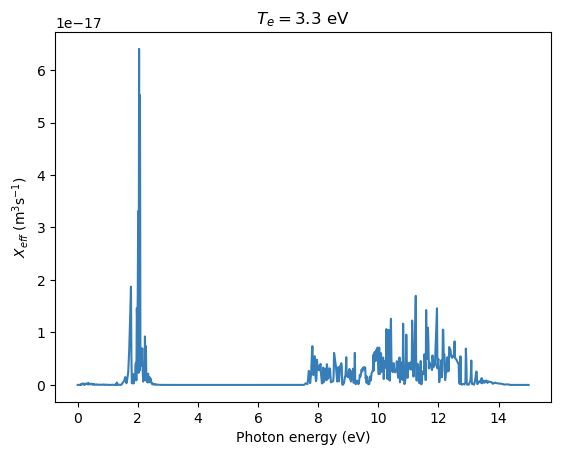}
         \caption{}
         \label{fig:Spectr_3.3}
     \end{subfigure}
     \hfill
     \begin{subfigure}[b]{0.49\textwidth}
         \centering
         \includegraphics[width=\textwidth]{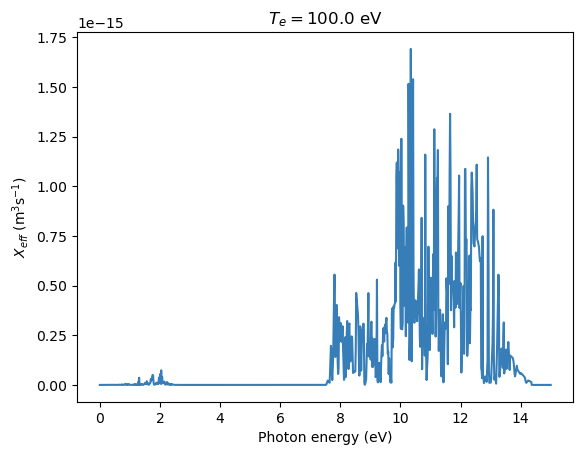}
         \caption{}
         \label{fig:Spectr_100}
     \end{subfigure}
        \caption{Emission spectra for different temperatures. The figures contain the total emission from all the different radiative transitions in the model. You can see that different photon energies dominate at different temperatures. Also note that $X_{eff}$ is much lower for $T_e = 3.3$ eV than for 100 eV. }
        \label{fig:Full_Spectra}
\end{figure}

One of the advantages of vibrationally resolving the electronically excited states is the ability to calculate the effective radiative emission rate coefficients $X_{eff}^{pq}$ for each transition from state p to q. This allows for the calculation of emission spectra as a function of photon energy, as well as the estimation of the total radiative power losses as a function of electron temperature and density. \autoref{fig:Full_Spectra} contains $X_{eff}$ as a function of photon energy, summed over all the radiative transitions in the model. The emission spectrum changes significantly as function of electron temperature. The relative emission (in terms of photons) from the $d^3\Pi_u$ to $a^3\Sigma_u$ transition, which is also called the Fulcher band, is stronger compared to that of the VUV emission (6-12 eV) at relatively low electron temperatures (2 eV). However, the absolute intensity of the Fulcher band is significantly reduced at such temperatures. The Fulcher band emission is important because it is often used as a diagnostic tool in spectroscopy as it emits in the visible regime ($\sim 2$ eV). More investigation into this particular transition is done in \autoref{sec:Fulcher}.

\begin{figure}[h]
     \centering
     \begin{subfigure}[b]{0.49\textwidth}
         \centering
         \includegraphics[width=\textwidth]{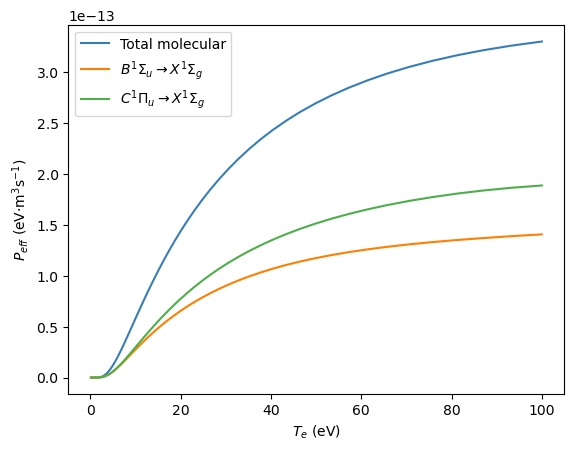}
         \caption{$P_{eff}$}
         \label{fig:Peff}
     \end{subfigure}
     \hfill
     \begin{subfigure}[b]{0.49\textwidth}
         \centering
         \includegraphics[width=\textwidth]{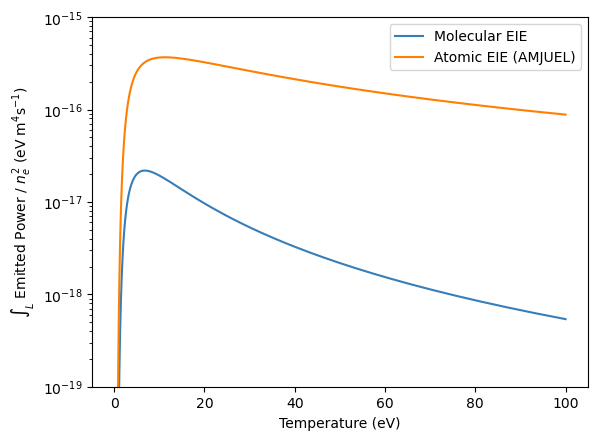}
         \caption{$f(T)P_{eff}$}
         \label{fig:Peff_int}
     \end{subfigure}
        \caption{(a) is a plot of the effective power emission coefficient as a function of temperature. The Lyman ($B^1\Sigma_u\rightarrow X^1\Sigma_g$) and Werner ($C^1\Pi_u\rightarrow X^1\Sigma_g$) transitions are plotted, as well as the total molecular $P_eff$, summed over all the transitions in the model. The Lyman and Werner transitions together make up over 95\% of all the emitted power in the system. (b) Is a plot of the calculated power emission, integrated over a line of sight $L$ (similar to \autoref{fig:rates_int}). The emitted power was calculated using scaling laws for the MAST-U Super-X divertor. Both the calculated molecular emission, and the expected atomic emission calculated using AMJUEL rates are given \cite{Eirene}. Different scaling laws were used for the molecular and atomic densities (see \autoref{sec:scalings}). }
        \label{fig:Peff+Peff_int}
\end{figure}

Using $X_{eff}$ as a function of energy, the power emission coefficients $P_{eff}$ can be calculated. \autoref{fig:Peff} shows the effective power emission coefficients for different temperatures. $P_{eff}$ increases as a function of electron temperature, because electron impact excitation of molecules is more likely at high temperatures. The Lyman ($B^1\Sigma_u\rightarrow X^1\Sigma_g$) and Werner ($C^1\Pi_u\rightarrow X^1\Sigma_g$) transitions make up over 95\% of the molecular radiation losses.

One of the goals of this report is to investigate the relative importance of molecular and atomic processes for radiative power dissipation. It is therefore interesting to compare the power emitted through Electron Impact Excitation (EIE) of molecules and atoms. This can be done by using the same toy model introduced in \autoref{sec:rates}: by using the scaling laws for $f_{\text{D}_2}(T)$ and $f_{\text{D}}(T)$, provided in \autoref{sec:scalings}. Because atomic excited states are not included in the CRM of \autoref{sec:model}, Eirene rates from the AMJUEL documentation are used for the atomic EIE \cite{Eirene}. The comparison between atomic and molecular power emission is given in \autoref{fig:Peff_int}. At temperatures above 0.3 eV, the molecular density decreases faster than the atomic density, so the atomic radiation dominates over the molecular radiation. At 100 eV there is a difference of more than two orders of magnitude. Although below 0.3 eV the power dissipated by molecular and atomic radiation is comparable, the radiated power decreases for both atomic and EIE at these temperatures. At $T_e=3$ eV the molecular contribution to the radiavive power losses through EIE is less than 1\%. Therefore the radiative power losses are unimportant at these temperatures.    

In short, this section has shown that the model of \autoref{sec:model} can be used to calculate emission spectra and estimate the radiative power loss from molecules. Even though different wavelengths dominate the spectrum at different temperatures, the combined power emitted by the Lyman and Werner transitions still proves to be over 95\% of the total molecular power losses. The comparison of \autoref{fig:Peff_int} between molecular and atomic radiation has shown that molecules play a significant role in the radiative power losses in the MAST-U Super-X divertor only for temperatures below approximately 0.6 eV. At these temperatures, however, there is hardly any radiation in the plasma, which means that the power dissipation through both molecular and atomic EIE is insignificant. This indicates that power losses due to \emph{molecular excited states} are likely not important in divertor plasmas. However, plasma-molecular interactions can lead to \emph{excited atoms} that radiate atomic emission. This can be significant \cite{verhaegh2023role}, but is out of scope of this work as it requires an integrated molecular and atomic CRM.

\subsection{Fulcher band emission spectra and distributions.}
\label{sec:Fulcher}
The emission from the transition from $d^3\Pi_u$ to $a^3\Sigma_g$, known as the Fulcher band emission, is often used as a diagnostic tool in spectroscopy. The Fulcher band emission can be used to experimentally determine the vibrational distribution in the $d^3\Pi_u$ state. The vibrational distribution of the $d^3\Pi_u$ state calculated using the model of \autoref{sec:model} can therefore be directly be compared to experiment. 

In \autoref{sec:Fulcher_background}, a method for modelling the vibrational distribution in the Fulcher upper state $d^3\Pi_u$ using Franck-Condon factors was given. Because the model of \autoref{sec:model} includes many different vibrationally resolved electronic states, the distribution in the upper Fulcher state can also be modelled directly from the CRM. To do this, it is necessary to include many different vibrationally resolved excited states, because the distribution in the $d^3\Pi_u$ state is highly influenced by the distributions of all the other states. The advantage of modelling all these states explicitly, in stead of mapping the ground state to the excited state, is that this method is more accurate.

\begin{figure}[H]
     \centering
     \begin{subfigure}[b]{0.49\textwidth}
         \centering
         \includegraphics[width=\textwidth]{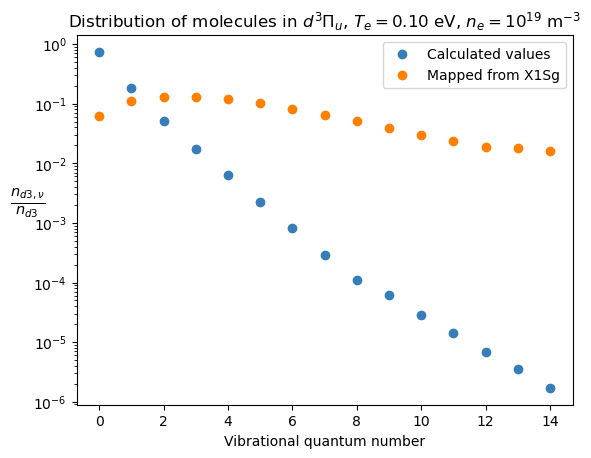}
         \caption{}
         \label{fig:d3_dist_low}
     \end{subfigure}
     \hfill
     \begin{subfigure}[b]{0.49\textwidth}
         \centering
         \includegraphics[width=\textwidth]{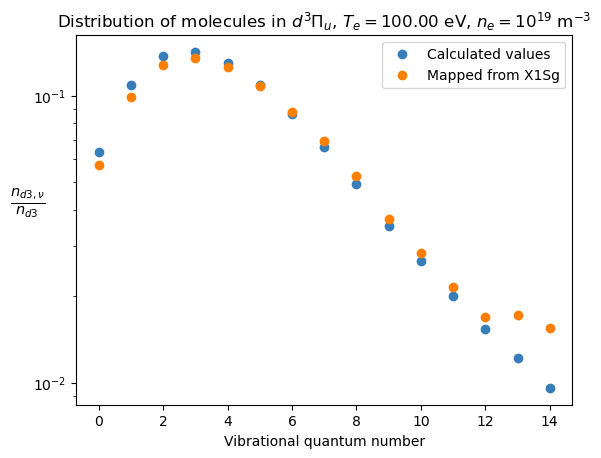}
         \caption{}
         \label{fig:d3_dist_high}
     \end{subfigure}
        \caption{Vibrational distribution in the $d^3\Pi_u$ state at different temperatures. $n_{d3,v}$ is the density of molecules in the $d^3\Pi_u$ state with vibrational quantum number $v$, and $n_{d3}$ is the total density of molecules in the $d^3\Pi_u$ state. The Franck-Condon mapping is quite close to the modeled distribution for higher temperatures, but deviates more at lower electron temperatures.}
        \label{fig:d3_dist}
\end{figure}

\autoref{fig:d3_dist} shows the vibrational distribution in the upper state for different temperatures. The figure contains both the distribution calculated using mapping from the ground state, and calculated directly from the model. The mapping closely resembles the calculated distribution for high temperatures, but deviates for lower temperatures. This indicates that there are processes at lower temperatures that are not taken into account by the mapping, and it is therefore necessary to model the excited states vibrationally resolved at these temperatures. However, as explained in section \autoref{sec:radiative_loss}, the Fulcher band intensity at such low temperatures is greatly diminished and the Fulcher band is generally too weak to be observed at temperatures below 1.5 eV \cite{Verhaegh_2023_fdens}.

\begin{figure}[h]
     \centering
     \begin{subfigure}[b]{0.49\textwidth}
         \centering
         \includegraphics[width=\textwidth]{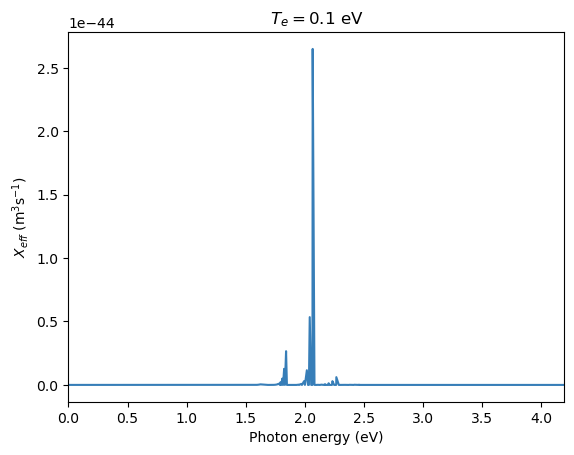}
         \caption{Fulcher band emission spectrum.}
         \label{fig:Spectr_Fulcher_01}
    \end{subfigure}
         \hfill
     \begin{subfigure}[b]{0.49\textwidth}
         \centering
         \includegraphics[width=\textwidth]{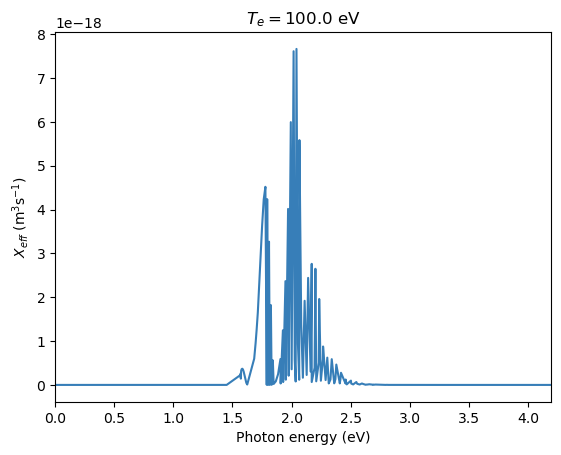}
         \caption{Fulcher band emission spectrum.}
         \label{fig:Spectr_Fulcher_100}
     \end{subfigure}
        \caption{Emission spectra for different temperatures. These spectra contain only the emission from the $d^3\Pi_u$ to $a^3\Sigma_g$ transition.}
        \label{fig:Spectra_Fulcher}
\end{figure}

In addition to generating spectra from all the different electronic transitions, it is possible to only look at the radiative transitions caused by the $d^3\Pi_u$ to $a^3\Sigma_g$ transition using the model of \autoref{sec:model}, thereby only observing the spectrum of the Fulcher band. \autoref{fig:Spectra_Fulcher} contains emission spectra from only the $d^3\Pi_u$ to $a^3\Sigma_g$ transition. These spectra can  directly be compared to experiment. Additionally, the generated spectra can help inferring information from the Fulcher band: the Fulcher band is such a mess of spectral lines that it's often difficult to infer which transition is from the Fulcher band and which transition isn't. Modelling the Fulcher band would greatly alleviate this. 
It is important to note here that to really make effective comparisons between the model and experimental results, rotational levels should also be included. The different energies of rotational states split the spectral lined of \autoref{fig:Spectra_Fulcher} into several different levels. The inclusion of rotational levels in the analysis of the model does not necessarily require a full rovibronic model: you can use an assumed rotational temperature to map the total vibronic intensities into estimates for the rovibronic intensities. 

The effective line emission coefficients $X_{eff}$ of \autoref{fig:Spectra_Fulcher} also allow for the estimation of the molecular density from measured Fulcher emission intensity, in combination with $n_e$ and $T_e$ measurements (see equation \eqref{eq:line_intensity}).  

\begin{figure}[h]
    \centering
    \includegraphics[width=0.5\textwidth]{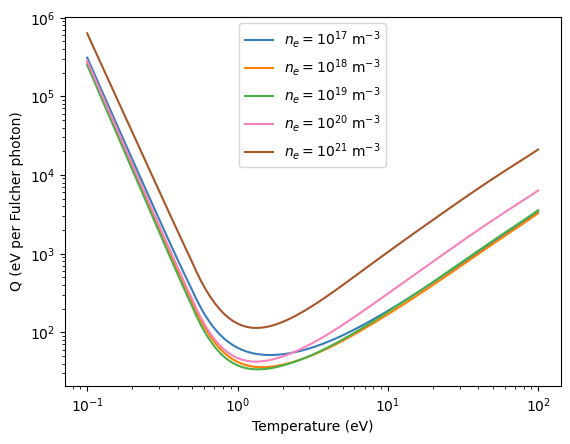}
    \caption{Q as a function of temperature for different electron densities.}
    \label{fig:Q}
\end{figure}

Because the number of photons emitted in the Fulcher band is often measured in experiments, it could be useful to have a relation between the number of photons from the Fulcher band and the total molecular radiative power loss. To do this, the factor $Q(T,n_e)$ was calculated.
\begin{equation}
    Q(T,n_e) = \frac{P_{eff}^{tot}}{X_{eff}^{d,tot}},
\end{equation}
where $X_{eff}^{d,tot}$ is the total emission rate coefficient in the Fulcher band, and $P_{eff}^{tot}$ is the total radiative power loss coefficient of the system. $Q$ is then the average energy lost to radiation per emitted photon in the Fulcher band. Since both $P_{eff}^{tot}$ and $X_{eff}^{d,tot}$ depend on temperature and density, so will $Q$. \autoref{fig:Q} contains $Q$ as a function of temperature for different electron densities. The radiative energy loss per Fulcher photon varies strongly for different temperatures. At low temperatures, there is very little excitation to $d^3\Pi_u$, so $Q$ is large. For high temperatures, the Lyman and Werner transitions (see \autoref{fig:E_diagram}) dominate, which again results in higher $Q$. 

\subsection{Vibrational-vibrational exchange.}
\label{sec:VVE}
One limitation of using equation \eqref{eq:matrices} is that only collisions between background species and molecules can be considered. However, especially in plasmas with high molecular and neutral atom density, heavy particle collisions might be important. In particular Vibrational-Vibrational Exchange (VVE) between two molecules colliding is expected to influence the vibrational distribution in the ground state \cite{Krasheninnikov1997}, and thereby the distribution in the upper Fulcher state. The reaction formula for VVE is given below. 

\begin{equation}
    \text{D}_2(X^1\Sigma_g,v)+\text{D}_2(X^1\Sigma_g,w+1)\rightarrow\text{D}_2(X^1\Sigma_g,v+1)+\text{D}_2(X^1\Sigma_g,w),
    \label{eq:VVE}
\end{equation}
$\text{D}_2(X^1\Sigma_g,v)$ are deuterium molecules in the electronic ground state with vibrational quantum number $v$, with $v>w$.

The problem is that these reactions involve two deuterium molecules, and are therefore quadratic in $n_0$. It is therefore not possible to linearize equation \eqref{eq:Rate_eq}, and write it as equation \eqref{eq:matrices}. It is possible, however to obtain an approximate steady-state distribution including VVE by using an iterative approach. 

Instead of solving the linear system of \eqref{eq:matrices} directly for steady state, the density distribution is evolved time dependently with some time step $\tau$. Equation \eqref{eq:matrices} is first solved to steady state. This yields an initial density profile $\mathbf{n}$ that is used to calculate the VVE reaction rates.

\begin{equation}
    R^{v,v+1}_{w+1,w} = n_vn_{w+1}k^{v,v+1}_{w+1,w},
\end{equation}

Where $R^{v,v+1}_{w+1,w}$ is the reaction rate (in cm$^{-3}$s$^{-1}$), $n_v$ is the density of $\text{D}_2(X^1\Sigma_g,v)$, and $k^{v,v+1}_{w+1,w}$ is the reaction rate coefficient (in cm$^3$s$^{-1}$) of equation \eqref{eq:VVE}. Using these rates, an additional source vector $\pmb{\Gamma}_\Delta$ is calculated, which represents the particle sources and sinks as a result of VVE. The components $\Gamma_{\Delta}^v$ are given by
\begin{equation}
    \Gamma_{\Delta,v} = -\sum_{w=0}^{v-1}R^{v,v+1}_{w+1,w} + \sum_{w=0}^{v-1}R^{v-1,v}_{w,w-1}.
\end{equation}
Using $\pmb{\Gamma}_\Delta$, the distribution is evolved up to $t=\tau$, using the following adapted equation. 
\begin{equation}
    \dot{\textbf{n}} = \textbf{M}\textbf{n}+\pmb{\Gamma}+\pmb{\Gamma}_\Delta.
\end{equation}
The densities $\mathbf{n}$ obtained from this equation can then be used to calculate the new $\pmb{\Gamma}_\Delta$, and repeat the process. This procedure is done until the densities converge to constant values. This way, a steady state distribution of the model including VVE is obtained. How the rate constants $k^{v,v+1}_{w+1,w}$ are calculated is given in \autoref{sec:VVE_rate}.

By doing these calculations, it was found that for reasonable values of the molecular density, i.e. $10^{19}$ m$^{-3}$, the gas temperature, i.e. 0.5 eV, and electron temperature, i.e. 1 eV, VVE does not play a significant role in the redistribution of particles in the ground state. Only once molecular density is increased to roughly $10^{24}$ m$^{-3}$ does VVE have any effect.


\subsection{Transport and equilibration time.}
\label{sec:transport}
Thus far, only steady-state solutions to equation \eqref{eq:matrices} ($\mathbf{\dot{n}}=0$) were considered. In reality, however, there will be transport of particles through the divertor, associated with some transport time $\tau_t$, which will prevent the particle distribution from equilibrating fully. The evolution of the density distribution as a function of time will depend on both density and temperature, and is characterized by the equilibration time $\tau_{eq}$. Transport may have a large effect on the density distribution and the reaction rates if $\tau_t<\tau_{eq}$. First, I use the various timescales to investigate in which regime (in terms of electron density/temperature) transport could play a role. 

\vspace{5mm}

\begin{figure}
    \centering
    \includegraphics[width=0.5\textwidth]{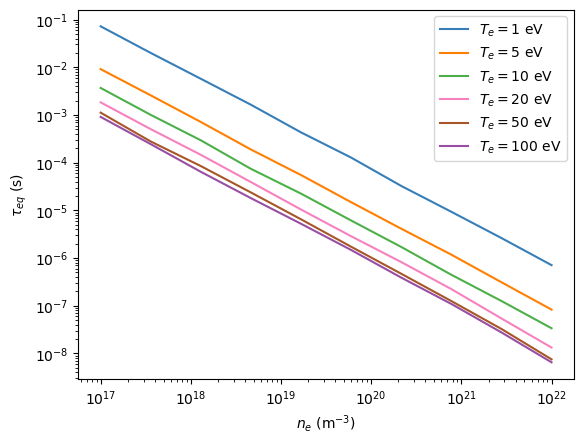}
    \caption{Equilibration time $\tau_{eq}$ as a function of electron density for different temperatures. $\tau_{eq}$ is found to be roughly inversely proportional to $n_e$ and highly dependent on the temperature. Depending on the temperature and density on the plasma, $\tau_{eq}$ may vary over many orders of magnitude.}
    \label{fig:teq}
\end{figure}

In order to quantify the convergence of the solution to a constant density distribution, I introduce the parameter $\epsilon$.
\begin{equation}
    \epsilon(t)=\frac{|\sum_i n_i^s-\sum_in_i(t)|}{\sum_in_i^s},
\end{equation}
where $n_i(t)$ is the density of species $i$ at time $t$, $n_i^s$ is the steady-state density of species $i$, and $\sum_i$ represents a sum over all the species in the system. For large $t$, $\epsilon$ will go to zero, as the densities approach the steady state distribution. For the purposes of this report, $\tau_{eq}$ is defined as the time it takes for $\epsilon(t)$ to permanently dip below $10^{-3}$. \autoref{fig:teq} contains $\tau_{eq}$ as a function of electron density for different temperatures. The equilibration time is found to be inversely proportional to $n_e$, and it varies over multiple orders of magnitude. For typical conditions in the MAST-U Super-X divertor, i.e. $T_e = 0.2$ eV and $n_e=10^{19}$ m$^{-3}$, the equilibration time is around 7 milliseconds.

To determine the effect of transport, the transport time $\tau_t$ must be estimated. The transport time is the time it takes for molecules in the plasma to travel large enough distances to experience significant differences in temperature. The transport time therefore depends on the thermal energy of the molecules and a characteristic length scale $l$. This length scale is the minimum distance over which significant temperature differences exist. $\tau_t$ is defined as the time it takes a molecule to traverse this length scale $l$ with ballistic motion. 
\begin{equation}
        \tau_t = \frac{l}{v} = l\sqrt{\frac{m_{\text{D}_2}}{2E}},
\end{equation}
where $v$ is the velocity, $E$ is the kinetic energy, and $m_{\text{D}_2}$ is the mass of the D$_2$ molecule. For a length $l=10$ cm, which is approximated from SOLPS-ITER simulations \cite{Myatra2023}, and energy of $E=0.5$ eV, which is approximated using Fulcher band analysis to estimate the rotational temperature (courtesy of N. Osborne), this gives a transport time of around 20 $\mu$s, which is much smaller than the previously calculated equilibration time at $T_e=0.2$ eV. Of course the exact value of the transport time will depend on the relevant length scale $l$ and molecular temperature, so this time scale may vary for different devices and discharges. 

Another important timescale for transport is the residence time of molecules in the plasma. This is the time associated with the destruction of molecules by ionizing or dissociative reactions. This time scale can be calculated using the following equation. 
\begin{equation}
    \tau_r = \frac{1}{n_e\langle\sigma v\rangle_{d}(T)},
    \label{eq:tres}
\end{equation}
where $\langle\sigma v\rangle_{d}(T)$ is the total effective rate coefficient for the destruction of deuterium molecules. $\tau_r$ is the time it takes for a molecule to be destroyed in one of the reactions of \autoref{fig:rates_n=19}. Based on this residence time, there are two different plasma regimes.

In the reaction limited regime $\tau_r<\tau_t$, molecules are destroyed faster than they can traverse the characteristic transport length $l$ - on average. The time molecules spend in a region with a certain electron temperature is limited by the frequency of the ionizing and dissociative reactions. Since transport is considered negligible for distances below $l$, its effects will be insignificant in the reaction limited regime. 

In the transport limited regime, $\tau_r>\tau_t$, molecules live long enough to travel a significant distance - on average. In this case, the time spent by a molecule in a region with a certain electron temperature is determined not by the reactions, but by the transport of the molecules. In this regime, particles will be able to travel farther than the characteristic length scale $l$, and therefore transport could be significant if the molecules do not equilibrate before they have travelled a significant difference ($\tau_{eq}>\tau_t$). 

\begin{figure}[h]
    \centering
    \includegraphics[width=0.5\textwidth]{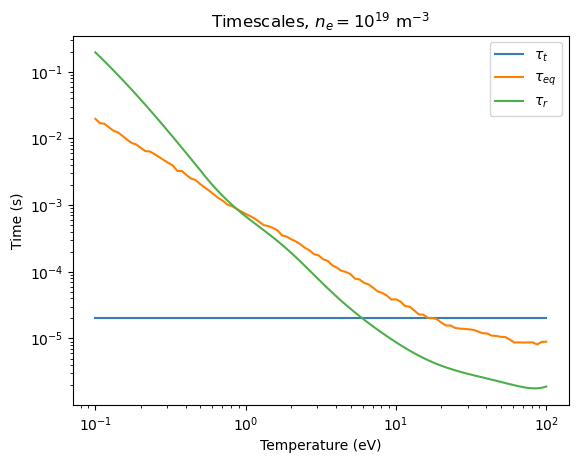}
    \caption{Relevant timescales $\tau_{eq}$, $\tau_t$ and $\tau_r$ as a function of temperature. $\tau_t\approx20$ $\mu$s, and was calculated using $E = 0.5$ eV and $l=10$ cm.}
    \label{fig:Timescales}
\end{figure}

\autoref{fig:Timescales} contains a comparison of the three timescales as a function of temperature. The cross over from the transport to reaction limited regime is at $T_e\approx6$ eV. We can assume therefore that no transport effects are present above this temperature. Below 6 eV, the transport time is smaller than the reaction time, so transport effects can be significant. Additionally, the equilibration time $\tau_{eq}$ below 6 eV is also larger than $\tau_t$, which also suggests that transport effects will play a role at these temperatures.  

\begin{figure}[h]
     \centering
     \begin{subfigure}[b]{0.49\textwidth}
         \centering
         \includegraphics[width=\textwidth]{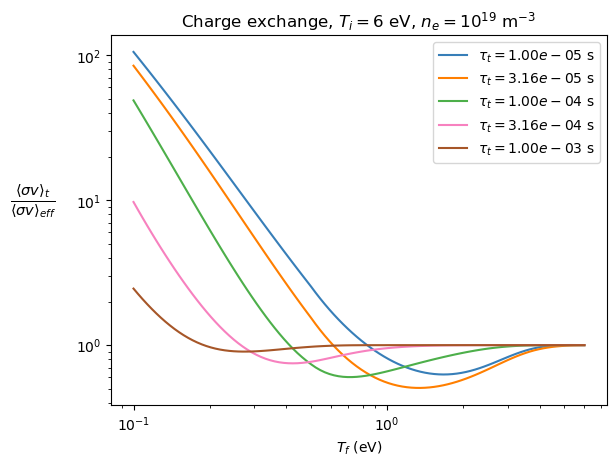}
         \caption{Molecular charge exchange}
         \label{fig:Transport_cx}
     \end{subfigure}
     \hfill
     \begin{subfigure}[b]{0.49\textwidth}
         \centering
         \includegraphics[width=\textwidth]{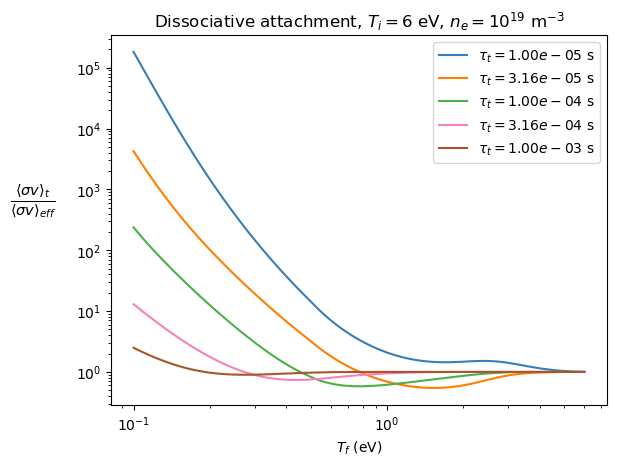}
         \caption{Dissociative attachment}
         \label{fig:Transport_da}
    \end{subfigure}
        \caption{Impact of transport on effective molecular charge exchange and dissociative attachment rates. The vertical axes show the effective rate with transport divided by the rates without transport.}
        \label{fig:Transport}
\end{figure}

The influence of transport on the effective reaction rates in the transport limited regime can be modelled using a simplified model. We assume an initial distribution $\mathbf{n}_i$, which is the same as the steady state distribution at some temperature $T_i\leq6$ eV. Equation \eqref{eq:matrices} is then solved up to the transport time $\tau_t$ at temperature $T_f$, resulting in the new distribution $\mathbf{n}_f$. This distribution is then used to calculate the effective rates in the region with temperature $T_f$. \autoref{fig:Transport} shows the effect of transport on the molecular charge exchange and dissociative attachment rates for different transport times $\tau_t$, in the case of particles travelling from an initial temperature $T_i=6$  eV, to a final temperature $T_f$. For sufficiently low $\tau_t$ and large temperature differences, transport phenomena can increase the molecular charge exchange rates in low temperature regions with up to a factor $10^2$. Transport may also decrease the amount of molecular charge exchange by a factor 2 for higher $T_f$. For dissociative attachment, transport may cause an increase of over $10^5$ for low transport times. This dramatic increase could lead to $D^-$ ions being more abundant than $D_2^+$ ions at low temperatures. If transport effects are included, MAR via $D^-$ might therefore exceed MAR via $D_2^+$ at these low temperatures. 

In conclusion, I have shown that for short transport times and large temperature differences, transport could increase the effective rate of molecular charge exchange up to a factor 100 in the divertor. This is a significant increase in the already underestimated molecular charge exchange rates. Additionally, it was shown that at electron temperatures below 6 eV, the plasma is in the transport limited regime. This indicates that transport between two regions with a large temperature difference is indeed possible. Therefore, transport could play a significant role at low temperatures in the MAST-U divertor. 
While this is an important conclusion for plasma edge modelling, the estimations made in this section are made based on a toy model. The characteristic transport length scale will of course depend on the (magnetic) geometry of the device, so the transport time will be different for each device. Additionally, in the calculations of $\tau_r$, we used the steady state effective rate coefficients of \autoref{fig:rates_n=19}. Because transport causes the actual distributions in the plasma to deviate from the steady state, this approximation is not entirely valid. Furthermore, the calculations of \autoref{fig:Transport} were done for an electron density of $10^{19}$ m$^{-3}$, which is typical for MAST-U, but reactors will probably operate at higher densities. Because both $\tau_t$ and $\tau_{eq}$ are shorter for higher densities, transport effects will be less severe.
Nevertheless, it was demonstrated that transport could have a significant effect on the molecular charge exchange rate at low temperatures under the right circumstances. It is therefore important to take transport effects into account for plasma edge simulations at sufficiently low temperatures, especially in the case of ADC's with high levels of detachment. 
The only way to quantitatively investigate the role of transport in divertor geometries numerically is to perform (at least partially) vibrationally resolved 2D or 3D simulations using models such as SOLPS. It might not be sufficient to simply calculate steady state distributions using a CRM, and then to use the resulting rates in larger 2D or 3D models, because the vibrational distributions in the plasma may not be the same as the steady state distributions due to transport effects. Vibrationally resolved plasma-exhaust simulations have been performed in the past \cite{Fantz2002,Wischmeier2005}. However, these were based on the (incorrect) H2VIBR rates and also, did not feature sufficiently deep detached conditions where molecular transport is expected to play a significant role. Hence, such interactions should be re-visited for ADC's.

\subsection{Plasma-wall interactions (PWI)}
\label{sec:PWI}
In the setup of the model in \autoref{sec:model}, the vibrational ground state of the D$_2$ molecule was used as a background species with constant density. It was thereby implicitly assumed that all the molecules entering the simulation volume are in the D$_2(X^1\Sigma_g, \nu=0)$ state, which is in general not the case for real plasmas. When the plasma is detached, neutral atoms in the divertor can recombine on the target surface, forming molecules. Depending on the energy of the atoms and the wall material, the molecules coming from the target will have a different vibrational distribution when entering the plasma. This affects the vibrational distribution in of the molecules in the plasma, thereby affecting the effective reaction rates and momentum, particle and energy losses. Rutgiliano et al. used molecular dynamics calculations to determine vibrational distributions of molecules produced by Eley-Rideal re-association on tungsten surfaces for different atomic kinetic energies \cite{Rutigliano_Cacciatore_2011}. These calculations were done for tungsten, which will be used in the divertor target of reactor class devices such as ITER \cite{ITER_2023}.

\begin{figure}
    \centering
    \includegraphics[width=0.5\textwidth]{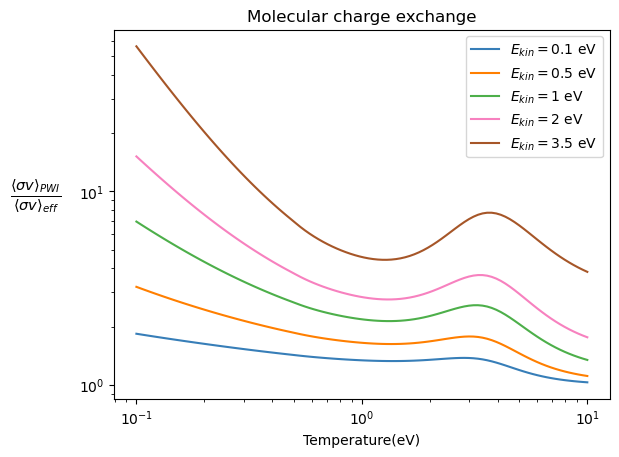}
    \caption{Effective charge exchange rate with PWI, divided by the rate without PWI. These results are based on the probability distributions of molecules coming from tungsten generated by Rutgiliano et al. at different atomic kinetic energies \cite{Rutigliano_Cacciatore_2011}. }
    \label{fig:PWI}
\end{figure}

The vibrational distribution of molecules coming from the target surface can be included in the model by adding an additional source vector $\Gamma_{PWI}$ to equation \eqref{eq:matrices}
\begin{equation}
    \dot{\textbf{n}} = \textbf{M}(T,n_i,...,n_m)\textbf{n}+\pmb{\Gamma}(T,n_i,...n_m) +\pmb{\Gamma}_{PWI}(E_{kin}).
    \label{eq:matrices_PWI}
\end{equation}
$E_{kin}$ is the kinetic energy of the deuterium atoms. Because the distribution of molecules coming from the tungsten surface is dependent on $E_{kin}$, so is the extra source vector $\Gamma_{PWI}$. Equation \eqref{eq:matrices_PWI} is then solved for steady state to determine the resulting distribution.  \autoref{fig:PWI} shows the effect of PWI on the effective charge exchange rate coefficient, under the assumption that all molecules entering the system come from Eley-Rideal re-association on the tungsten surface. Depending on the kinetic energy of the atoms, and the electron temperature of the plasma, the re-association of deuterium atoms on tungsten may increase the charge exchange rate up to a factor of 60. For typical conditions in MAST-U, the temperature of the atoms is around 1 eV, so the influence of PWI would be less, but still significant. At high temperatures, the effect of PWI decreases, due to the higher reaction rates.

In conclusion, this section has shown that plasma-wall interactions can play a significant role in the vibrational distribution of the molecules in the plasma, particularly at low temperatures. Therefore, to model divertor plasmas accurately, it is necessary to know the interaction dynamics of neutral atoms with the target. The results of \autoref{fig:PWI} are calculated for the case that all molecules that enter the plasma are created by re-association on tungsten, which is not the case for real divertor plasmas. Additionally, it assumes that all the surface re-associated molecules arise from the Elay-Rideal process. In reality, there is a host of different plasma-surface interactions, including reflection from the wall before re-association occurs and including the formation of surface layers on the walls \cite{Wischmeier2005}. It should also be noted that MAST-U has a carbon divertor and plasma-surface interactions on carbon are significantly different from that on tungsten: reflection from the wall is more likely for tungsten, whereas adsorption on the wall occurs for carbon due to the chemical affinity between carbon and hydrogen, which could modify the vibrational distribution of molecules being released from the wall \cite{Wischmeier2005}.

Regardless, I have shown that the vibrational distribution of molecules entering the simulation volume can have a significant impact on the steady state distributions. In order to do accurate simulations of detached divertor plasmas, we therefore need an in-depth understanding of the physics of D$_2$ formation, both through re-association on the surface and otherwise. 

%% file: 4_Conclusions.tex
\newpage
\section{Conclusions}
\label{sec:Conclusions}
In the introduction, three different topics, each with several research questions, have been posited, which were investigated in the results section. In this section, these questions are revisited and the conclusions drawn from the research are presented.

\vspace{5mm}

\textbf{1. What does the new CRM say about the impact of plasma-molecular interactions in exhaust processes?}

\emph{
\begin{itemize}
    \item[1.1] How do effective rates (calculated using the new CRM) compare to default rates tabulated in Eirene, and how significant would these rate differences be for the effective ion sources/sinks in the plasma?
\end{itemize}
}

In \autoref{sec:rates} it was demonstrated that the inaccuracies in the estimation of molecular charge exchange made by Eirene lead to underestimation of the of plasma-molecule interactions of multiple orders of magnitude at low temperatures. \autoref{fig:MA_plus_n=19} showed that the effective rate coefficients for MAR and MAD calculated using the model of \autoref{sec:model} are significantly larger at low temperatures than the AMJUEL rates. When a toy model is used to model the ion sources and sinks using scaling laws for the molecular density as function of temperature for MAST-U, the MAR rate is roughly constant at low temperatures \cite{Verhaegh_2023_fdens}. When compared to calculations using AMJUEL data for EIR, MAR contributes significantly to the ion total ion sinks, (see \autoref{fig:ions}). Rates obtained from AMJUEL suggest that MAR is insignificant at low temperatures, while the model of \autoref{sec:model} shows significant MAR even at temperatures below 1 eV. However, for higher densities it was shown that EIR increases, while MAR is mostly unaffected. At $n_e=10^{19}$ m$^{-3}$, MAR is larger than EIR at temperatures between 0.5 and 3.5 eV, while at $n_e=10^{20}$ m$^{-3}$, MAR exceeds EIR between 0.4 and 2 eV. MAR is therefore predicted to be less important for the total effective ion sinks at higher electron density. Note, however, that future reactors are expected to operate at higher temperatures than MAST-U. If the lowest target temperatures in a future reactor are around 1 eV, MAR might still be more important than EIR, even at higher densities. The model thus indicates a significant effect of molecular interactions on ion sinks (MAR) and hydrogenic power losses in the plasma. This prediction agrees well with experiments done on MAST-U, which have shown elevated levels of MAR in highly detached divertor plasmas \cite{verhaegh2023role}. 

\emph{
\begin{itemize}
    \item[1.2] What is the role of electronically excited molecules in the plasma?
\end{itemize}
}

In \autoref{sec:exc} the impact of electronically excited molecules was investigated. \autoref{fig:vibr_dist_unr+no} shows that the presence of electronically excited states in the model greatly influences the vibrational distribution of the molecules in the ground state above $T_e=2$ eV, because, at these temperatures, electronic excitation through electron collisions becomes more significant. By comparing \autoref{fig:vibr_dist_full} and \autoref{fig:vibr_dist_unr}, it was demonstrated that the inclusion of highly excited molecular states, as well as vibrationally resolving the lower excited states has only a limited effect on the vibrational distribution. \autoref{fig:cx_excited_states} also shows that the inclusion of electronically excited states makes a difference above 2 eV, but including many vibrationally excited has little to no effect on the effective molecular charge exchange rate. Furthermore, the number of electronically excited molecules was found to be small compared to the ground state ($<1\%$), which indicates that the vibrational distribution in the excited states will have negligible effect on the plasma chemistry. It can therefore be concluded that redistribution of molecules in the ground state through transitions to and from electronically excited states is important, but it is probably unnecessary to add lots of vibrationally resolved excited states to the model. It is likely sufficient to add only the first few electronic states unresolved. Note that for the calculation of emission spectra, the vibrational distributions in the excited states are important, and must therefore be included in the model if we want to use it to do spectroscopic analysis. 

\emph{
\begin{itemize}
    \item[1.3] What is the relative importance of radiative power dissipation caused by electronically excited molecules compared to excited atoms?
\end{itemize}
}

Analysis of emission spectra from the model in \autoref{sec:radiative_loss} has showed that the dominant molecular emission wavelengths are different for different electron temperatures. Analysis of the $D^*$ radiative cooling rate has demonstrated that over 95\% of the $D^*$ radiative power loss is generated by the Werner and Lyman transitions. Using a toy model for the atomic and molecular electron-impact excitation radiation, \autoref{fig:Peff_int} demonstrates that molecular radiation can be a significant contributor to power loss compared to atomic radiation at $T_e<1$ eV. Note that at these temperatures, EIE decreases rapidly for both atoms and molecules. This is also confirmed by \autoref{sec:exc}, which shows that the amount of electronically excited molecules is negligible at temperatures below 1 eV. At these temperatures, radiative power losses are therefore unimportant.

\vspace{5mm}

In conclusion, the model of \autoref{sec:model} yields effective reaction rates that have a better correspondence to experiment than the commonly used Eirene rates. It was shown that MAR can be a significant contributor to the total recombination processes in the divertor. EIE of molecules is shown to only really be significant at temperatures below $0.3$ eV, where there is almost no radiation, so molecules are not expected to have a large impact on radiative power dissipation in divertor plasmas. In general, electronically excited states are expected to only have an impact on the plasma processes through redistribution of molecules in the ground state. 

\vspace{5mm}

\textbf{2. Can the new CRM be used to improve experimental data analysis of plasma-molecular interactions?}

\emph{
\begin{itemize}
    \item[2.1] Under which conditions is the simple model involving Franck-Condon factors valid?
\end{itemize}
}

The distribution in the upper Fulcher state is compared to a mapping from the ground state, in \autoref{fig:d3_dist}. The mapping was calculated using Franck-Condon factors as weights to map the distribution in the electronic ground state to the upper state. It was found that the mapping is quite close to distribution calculated directly from the model for high temperatures. For low temperatures, however, the vibrational distribution deviates from the mapping quite significantly. This indicates that it is not possible to obtain the vibrational distribution in the ground state by measuring the distribution in the upper Fulcher state, at low temperatures. The mapping of \eqref{eq:mapping} proves to be inaccurate at low temperatures, so the measurement of the vibrational distribution in the ground has to be done directly, which could be achieved with active laser spectroscopy (VUV-LIF \cite{Vankan2004}). Additionally, the Fulcher emission may be insufficient at low electron temperatures to facilitate an analysis of the vibrational distribution using visible emission spectroscopy.
It is important to mention here that the $d^3\Pi_u$ state is the highest electronic state included in the model of \autoref{sec:model}. In reality, however, there are higher electronically excited states of the deuterium molecule, as can be seen in \autoref{fig:E_diagram}. As demonstrated in \autoref{sec:exc}, these higher states probably do not have a significant influence on the effective rates of \autoref{sec:rates}. They could, however, be important for the vibrational distribution in the upper Fulcher state. 

\emph{
\begin{itemize}
    \item[2.2] Can the CRM of this report be used to generate emission spectra that can be used to improve analysis of experimental data?
\end{itemize}
}

The explicitly calculated vibrational distribution in the upper Fulcher state was used to calculate effective radiative emission coefficients for the Fulcher band emission. By measuring the Fulcher intensity and other plasma parameters such as electron density and temperature, the emission rate coefficients can be used to estimate the molecular density and molecular radiative power dissipation in real divertor plasmas. The calculated emission spectra in \autoref{fig:Spectra_Fulcher} can also be compared to experiment, if rotational levels are added to the analysis.

\vspace{5mm}

The model of \autoref{sec:model} has therefore somewhat validated the use of the Franck-Condon mapping of equation \eqref{eq:mapping}, at least at higher temperatures. \autoref{fig:Spectra_Fulcher} shows that the CRM of \autoref{sec:model} can be used to generate emission spectra from the Fulcher transition. These spectra could be used to improve analysis of experimentally measured emission spectra. Note that inclusion of rotationally resolved states will probably be necessary to fully make use of the spectra from the model. This can likely be done quite easily by assuming a Bolzmann distribution for the rotational distribution, which a fixed rotational temperature in each vibrational state. Therefore, due to the vibrationally resolved excited states of in the model of \autoref{sec:model}, it could be a valuable asset to improve spectroscopic experimental data analysis of plasma-molecular interactions. 

\vspace{5mm}

\textbf{3. Can processes that prevent using effective rates in codes be important?}

\emph{
\begin{itemize}
    \item[3.1] Under which conditions is there a significant effect of vibrational-vibrational exchange on the vibrational distribution; and under which conditions can this significantly impact plasma-molecular reaction rates?
\end{itemize} 
}

In \autoref{sec:VVE}, the role of Vibrational-Vibrational Exchange (VVE) between molecules was investigated using an iterative method to arrive at an approximate solution to equation \eqref{eq:Rate_eq}. It was found that using an approximate relation for the VVE rate coefficients by Matveyev and Shikalov from 1995, VVE does not contribute significantly to the redistribution of molecules in the electronic ground state for conditions relevant to the MAST-U Super-X divertor. Using the approach explained in \autoref{sec:VVE}, VVE only has significant effects for molecular densities above $10^{24}$ m$^{-3}$. This is possibly due to the low molecular densities and high electron temperatures in fusion plasmas, which cause other redistribution mechanisms such as electron collisions and excitation followed by radiative decay to be much more important. VVE is therefore unlikely to have a significant impact on the molecular processes in MAST-U. 

\emph{
\begin{itemize}
    \item[3.2] Under which conditions can transport have a significant impact on the vibrational distribution of the molecules as well as on the MAR ion sink?
\end{itemize} 
}

Section \ref{sec:transport} introduces three timescales relevant for transport: the transport time $\tau_t$, related to the characteristic length scale and thermal energy of the molecules; the reaction time $\tau_r$, which is defined as the time it takes for a molecule to be destroyed by charge exchange, dissociation, ionization or dissociative attachment; and the equilibration time $\tau_{eq}$, related to the time it takes for a particle distribution to converge to the steady state solution at a certain temperature and density. It was found that transport is unlikely to have significant effects at temperatures above $T_e=6$ eV at $n_e = 10^{19}$, because for higher temperatures $\tau_r<\tau_t$. Below this temperature, however transport can have a significant effect on the molecular charge exchange rate. For a distribution travelling from an initial electron temperature $T_i=6$ eV to some final temperature $T_f<T_i$, it was found that the molecular charge exchange rate may increase with up to a factor of $10^2$. It was discovered that the molecular charge exchange rate can also decrease for certain final temperatures. This is probably due to the fact that the molecular charge exchange rates calculated by Ichihara et al. have a resonance at $\nu=4$. Therefore, the vibrational distribution at $T_e=6$ eV is less favorable for molecular charge exchange than at approximately 1 eV. This can also be seen from \autoref{fig:rates_n=19}, and from the fact that the fractional abundance of states $\nu=4$ to $\nu=9$ is larger at 1 eV than at 6 eV. The calculated effects of transport on the dissociative attachment rates are even more pronounced than for molecular charge exchange, with an increase of a factor of up to $10^5$. Therefore, if transport effects are dominant, the D$^-$ concentration in the plasma may exceed the D$_2^+$ concentration, wich could lead to MAR via D$^-$ dominating over MAR via D$_2^+$. This further empowers the claim that negative ions may in fact play a significant role in deuterium plasmas. In the low temperature conditions of the Super-X divertor in MAST-U, transport is thus expected to have a significant influence on the vibrational distribution, and thereby the effective rates of molecular processes. 

\emph{
\begin{itemize}
    \item[3.3] Under which conditions can the influx of vibrationally excited molecules from the wall after plasma wall interactions have a significant impact on the MAR ion sink?
\end{itemize} 
}

Rutigliano et al. demonstrated in 2011 that hydrogen atoms that recombine on a tungsten surface will inject molecules into the plasma with a certain vibrational distribution \cite{Rutigliano_Cacciatore_2011}. In \autoref{sec:PWI}, the vibrational distributions from Rutigliano's 2011 paper were used to calculate source vectors for equation \eqref{eq:matrices}, to investigate the potential impact of plasma-wall interactions (PWI) on the effective molecular charge exchange rate. It was found that for high thermal energy of the atoms and low electron temperature, PWI can increase the effective charge exchange rate with a factor of around 60, compared to the case without PWI. The data of \autoref{fig:PWI} is calculated under the assumption that all molecules entering the simulation come from the re-association on the tungsten surface, which will in general not be true for real plasmas. Additionally, MAST-U has a carbon wall and the vibrational distribution of the molecules released from the carbon wall could deviate significantly from that of tungsten due to the chemical affinity between carbon and hydrogen. Nevertheless, the results of \autoref{fig:PWI} show that the vibrational distribution of the molecules entering the plasma at low temperatures has a significant influence on the molecular processes. To model these processes accurately, then, detailed understanding of the processes that create molecules in the divertor, and the vibrational distribution they produce are needed.

\vspace{5mm}

Ultimately, it was shown that at least in the case of transport effects and PWI, processes that prevent the use of effective rates in codes can be important under realistic divertor plasma conditions. This is because these processes can alter the vibrational distribution of deuterium molecules in the ground state, which in turn affects the effective reaction rates for plasma-molecular interactions. This leads to the conclusion that to model the effect of molecules in divertor plasmas accurately, a deep understanding of processes such as transport and PWI is needed. The only way to truly investigate the impact of these processes quantitatively, vibrationally resolved 2D plasma edge models, that take both PWI and the geometry of the divertor into account are needed. 

\vspace{5mm}

In conclusion, in this report it has been demonstrated that molecular processes can be  relevant in highly detached plasmas, and that these processes are possibly affected by processes such as transport and PWI. This leads to the conclusion that, to model highly detached divertor plasmas, it may not be sufficient to use CRM's to generate effective reaction rates as a function of temperature and density in steady state conditions, and use these rates as input for SOLPS simulations. This is because in general, the vibrational distributions in the divertor may not correspond to the steady state solutions of CRM's, but could rather depend on the specifics of the (magnetic) geometry of the divertor and the processes that create molecules. This means that if we want to be able to predict heat and particle fluxes in the divertor of future reactors such as DEMO, which will need high levels of detachment, it may be necessary to develop more sophisticated plasma edge simulations that explicitly take the vibrational distribution of molecules into account, as well as the different processes that determine this distribution \cite{Fil_2022}.

%% file: 5_Discussion.tex
\section{Discussion}
In this section, the results of \autoref{sec:Results} and the conclusions of \autoref{sec:Conclusions} are discussed in some more detail. 




\subsection{Model restrictions}
Even though the model of \autoref{sec:model} is closer to experiment than the data in Eirene documentation, it is not without its faults. Firstly, although most rates are taken from calculations for D$_2$, this data is not available for all relevant reactions. For example, electron impact excitation of electronically excited molecules is only available for H$_2$, and not vibrationally resolved. To include them in the model anyway, Franck-Condon factors were used to estimate the excitation rates between two vibrationally resolved excited states. This is, however, not entirely appropriate, since the Franck-Condon factors apply only to transitions by emission or absorption of a photon, not by electron collisions. It is unclear how much use of hydrogen rates and Franck-Condon factors in the calculation of electron impact excitation rates influences the results of the model, and to investigate the effect, the rates for these transitions need to be determined for vibrationally resolved deuterium. 

The molecular charge exchange rates of Ichihara are for hydrogen, and therefore also not entirely applicable for deuterium. Even though the temperature rescaling in this model is done correctly, the specific resonances for this reaction are different for deuterium compared to hydrogen, which could affect the effective rates. Additionally, the rate data from Ichihara is only given for temperatures from 0.1 to 5 eV. To obtain the rates outside this range, the data was extrapolated. Again it is unclear how these approximations will affect the effective rates exactly, and to test it requires rate coefficients of molecular charge exchange specifically for deuterium. Some other similar approximations are used in the model, all of which are listed in \autoref{sec:rate_explanation}.

Another restriction in the current model concerns the reactions that deplete $D_2^+$ and $D^-$ ions, which strongly impact the effective MAR, MAI and MAD rates. The rates used for these reactions are still from tabulated AMJUEL data, and should likely also be re-evaluated. 

The model used does not resolve the excited levels of the hydrogen atoms (i.e., it is a molecular CRM, not an atomic CRM). This may lead to overestimations of the electron-impact dissociation rate at high electron densities ($>10^{21}$ m$^{-3}$), as some of such dissociation events may produce highly excited atoms that are more easily re-ionised from electron collisions \cite{Sawada1995}.


\subsection{Possible improvements of the current model}
To improve upon the model of \autoref{sec:model}, the first step could be to rectify the restrictions of the previous section by calculating rates for all the different transitions in the model specifically for deuterium. Additionally, the model may be improved by adding more electronically excited states for the molecules. As previously stated in \autoref{sec:Conclusions}, the vibrational distribution the Fulcher state could be affected by redistribution via transitions to and from higher electronic states. This could have an effect on the Fulcher emission spectrum and intensity. The problem with including these states is that very little data is available for reactions involving these states. So to include these, additional ab initio quantum mechanical calculations need to be performed to include these states, or simplified approximations must be made to 'rescale' data from lower electronic excited levels to higher electronic excited levels as has been used previously \cite{Sawada1995}. 

\begin{figure}
    \centering
    \includegraphics[width=0.5\textwidth]{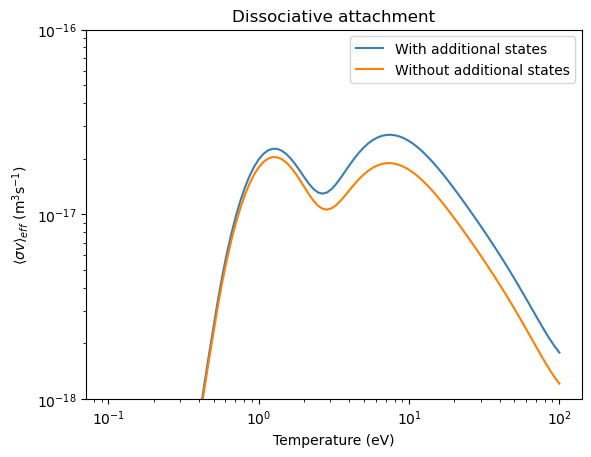}
    \caption{Comparison of dissociative attachment rates. The orange curve is simply the dissociative attachment rate calculated using the model of \autoref{sec:model}, and the blue curve takes higher vibrational states into account. }
    \label{fig:DA_states}
\end{figure}

In the model of \autoref{sec:model}, it is explained that the electronic states are vibrationally resolved up to $\nu=14$. However, molecular deuterium can also exist for higher vibrational states. These states could have an impact on the effective reaction rates and vibrational distributions in the system. These states also give rise to extra radiative transitions, so the spectra that result from a model with more vibrational states could be quite different. These states could be included in the model explicitly by adding more species to the CRM, but this is quite computationally expensive. In stead, it might also be possible to assume a Bolzmann distribution for the states higher than $\nu=14$ and use this to calculate the effective rates. I have used this method to estimate the effect of higher vibrational states on the dissociative attachment rates. The results are given in \autoref{fig:DA_states}. In this figure, I have included the vibrational states up to $\nu=20$ in the ground state and $\nu=50$ in the $B^1\Sigma_u$ state by extrapolating the vibrational distribution from the model for values above $\nu=14$. The difference between the two rates illustrates that it might be necessary to include higher vibrational states in the model, either by explicitly adding them, or by inferring them using a Bolzmann distribution.

In addition to vibrational excitation, molecular deuterium can also be excited rotationally. These levels were not included in the model of \autoref{sec:model}, because unlike to the vibrational state, the rotational state is not expected to significantly influence the reaction rates. 

However, the emission spectra are influenced by rotational excitation. If these levels are included, it might be possible to generate more accurate emission spectra to compare with experimental measurements. These levels could be included explicitly by going to a full rovibronic model, but this would be computationally expensive, and probably unnecessary. We could also assume the rotational states are distributed according to a Bolzmann distribution with a certain rotational temperature. This temperature could even vary for different vibrational states. Going to a full rovibronic model only makes sense if the electron and ion collisions that drive the rotational states are expected to impact the reaction rates, or the electronic/vibrational distribution. If this is the case, it is not sufficient to calculate a vibrational distribution and use the assumption of a Bolzmann distribution to calculate the rotational states. There are currently no cross sections or rates available for rotationally resolved molecules, so these would have to be obtained also via quantum mechanical calculations, if we want to go to a full rovibronic model, or simplified approximations to 'rescale' rates must be made \cite{Sawada2016}. 

In the model of \autoref{sec:model}, the atomic states of deuterium are not included, which is why AMJUEL data is used for the comparisons between atomic and molecular processes. The densities of atomic excited states in the plasma could affect the overall plasma dynamics. Therefore, a next step would be to include atoms in the model. 

Furthermore, since $D_2^+$ also consists of 2 nuclei, it can also be vibrationally and electronically excited. Like for molecules, the vibronic state of $D_2^2$ may influence the reaction rates, which is not accounted for in the model of \autoref{sec:model}. Vibrationally resolved excitation and dissociation cross sections for $D_2^+$ ions are given on the MCCCDB website, and could be included in the model. 

\subsection{Implications of the current model and suggestions for exhaust modelling.}

Unfortunately, even if all relevant processes are included in the CRM with perfectly accurate rates, we will still be unable to calculate the effects of transport and PWI without models that take spatial geometries into account. As stated in the introduction, the ultimate goal of the research into this part of plasma physics is to develop plasma modelling tools that are able to accurately simulate divertor physics, even in highly detached plasmas. This means that vibrationally resolved setups will have to be used, either by altering existing tools such as SOLPS, or by developing entirely new simulation tools. 

However, it might not be necessary to include all vibrationally and electronically resolved states in these simulations. As demonstrated by Greenland et al. in 2001, there are sophisticated methods for reducing the number of species that need to be explicitly tracked in the simulation \cite{Greenland_Molecules}. This is done by separating the species into so-called P and Q states. The P states are then tracked explicitly, while the densities of the Q states can be determined using a simple relation from the densities of the P states. This method reduces the system of equation \eqref{eq:matrices} from an $N\times N$, to an $N_p \times N_p$ system, where $N$ is the total number of states in the system, and $N_p$ is the number of P states. Greenland provides in their 2001 paper an algorithm for automating the choosing of the P and Q states \cite{Greenland_Molecules}. If implemented correctly, this algorithm could significantly reduce the required computational power. Unfortunately, we were not able to properly implement Greenland's method in the allotted time for this research project. This was mainly due to the fact that the method appears to be highly sensitive to numerical instabilities. In theory, however, it should be possible to apply the method to our model by cleaning up the matrix and altering any rates that might cause numerical instabilities. 

I therefore recommend a thorough revision of the commonly used molecular reaction rates tabulated in Eirene, as well as more extensive investigation into the possible effects of transport and plasma-wall interactions on the vibrational distribution of the molecules.

%% file: Acknowledgements.tex
\section*{Acknowledgements}

The results are obtained with the help of the EIRENE package (see www.eirene.de) including the related code, data and tools \cite{Eirene}. This work has received support from EPSRC Grants EP/T012250/1 and EP/N023846/1. This work has been carried out within the framework of the EUROfusion Consortium, partially funded by the European Union via the Euratom Research and Training Programme (Grant Agreement No 101052200 — EUROfusion). Views and opinions expressed are however those of the author(s) only and do not necessarily reflect those of the European Union or the European Commission. Neither the European Union nor the European Commission can be held responsible for them.

The CRM's in this report are generated using CRUMPET \cite{HOLM2021100982}. The git repository of CRUMPET can be found at https://github.com/holm10/CRUMPET. 

Finally, this work was done with the support of both the United Kingdom Atomic Energy Authority (UKAEA) and FuseNet.

%% file: Appendix_A.tex
\section{List of reactions}
\label{sec:reaction_list}

\begin{itemize}
    \item Molecular charge exchange from $X^1\Sigma_g$: $\text{D}^++\text{D}_2(n,\nu)\rightarrow \text{D} + \text{D}_2^+$ \cite{Ichihara_2000};
    \item Dissociative attachment from $X^1\Sigma_g$ and $B^1\Sigma_u$:  $e+\text{D}_2(n,\nu)\rightarrow \text{D}^-+\text{D}$ \cite{Laporta_2021};
    \item Vibrational transitions (collisional excitation) in $X^1\Sigma_g$ and $B^1\Sigma_u$: $e+\text{D}_2(n,\nu=i)\rightarrow e+\text{D}_2(n,\nu=j)$ \cite{Laporta_2021};
    \item Dissociative excitation from $X^1\Sigma_g$: $e+\text{D}_2^+\rightarrow e + \text{D}^+ + \text{D}$ \cite{MCCCDB_exc_X1};
    \item Ionization from $X^1\Sigma_g$, $B^1\Sigma_u$, $C^1\Pi_u$, $EF^1\Sigma_g$, $a^3\Sigma_g$, $c^3\Pi_u$ and $d^3\Pi_u$: $e+\text{D}_2(n,\nu)\rightarrow 2e+\text{D}_2^+$ \cite{MCCCDB_Ionization};
    \item Collisional excitation and deexcitation:  $e+\text{D}_2(n=A,\nu=i)\rightarrow e+\text{D}_2(n=B,\nu=j)$. The following electronic transitions are included.  
    \begin{itemize}
        \item $X^1\Sigma_g \leftrightarrow B^1\Sigma_u$ \cite{MCCCDB_exc_X1};
        \item $X^1\Sigma_g \leftrightarrow C^1\Pi_u$ \cite{MCCCDB_exc_X1};
        \item $X^1\Sigma_g \leftrightarrow EF^1\Sigma_g$ \cite{MCCCDB_exc_X1};
        \item $X^1\Sigma_g \leftrightarrow a^3\Sigma_g$ \cite{MCCCDB_exc_X1};
        \item $X^1\Sigma_g \leftrightarrow c^3\Pi_u$ \cite{MCCCDB_exc_X1};
        \item $X^1\Sigma_g \leftrightarrow d^3\Pi_u$ \cite{MCCCDB_exc_X1};
        \item $B^1\Sigma_u \leftrightarrow C^1\Pi_u$ \cite{MCCCDB_exc_exc,FANTZ2006853};
        \item $B^1\Sigma_u \leftrightarrow EF^1\Sigma_g$ \cite{MCCCDB_exc_exc,FANTZ2006853};
        \item $C^1\Pi_u \leftrightarrow EF^1\Sigma_g$ \cite{MCCCDB_exc_exc,FANTZ2006853};
        \item $a^3\Sigma_g \leftrightarrow d^3\Pi_u$ \cite{MCCCDB_exc_exc,FANTZ2006853};
        \item $c^3\Pi_g \leftrightarrow d^3\Pi_u$ \cite{MCCCDB_exc_exc,FANTZ2006853};
    \end{itemize}
    \item Radiative deexcitation: $\text{D}_2(n=A,\nu=i)\rightarrow\text{D}_2(n=B,\nu=j)+\hbar\omega$ \cite{FANTZ2006853};
    \begin{itemize}
        \item $B^1\Sigma_u \rightarrow X^1\Sigma_g$;
        \item $C^1\Pi_u \rightarrow X^1\Sigma_g$;
        \item $EF^1\Sigma_g \rightarrow X^1\Sigma_g$;
        \item $EF^1\Sigma_g \leftrightarrow B^1\Sigma_u$;
        \item $EF^1\Sigma_g \leftrightarrow C^1\Pi_u$;
        \item $a^3\Sigma_g \rightarrow b^3\Sigma_u$ (dissociation);
        \item $a^3\Sigma_g \leftrightarrow c^3\Pi_u$;
        \item $d^3\Pi_u \rightarrow a^3\Sigma_g$
    \end{itemize}
    \item Reactions with D$_2^+$ \cite{Eirene}:
    \begin{itemize}
        \item $e+\text{D}_2^+\rightarrow 2\text{D}$;
        \item $e+\text{D}_2^+\rightarrow e + \text{D}^++\text{D}$;
        \item $e+\text{D}_2^+\rightarrow 2e+2\text{D}^+$;
    \end{itemize}
    \item Reactions with D$^-$ \cite{Eirene}:
    \begin{itemize}
        \item $\text{D}^++\text{D}^-\rightarrow 2\text{D}$;
        \item $\text{D}^++\text{D}^-\rightarrow e + \text{D}^++\text{D}$;
    \end{itemize}
\end{itemize}

%% file: Appendix_B.tex
\section{Explanation of rate calculations}
\label{sec:rate_explanation}
This section contains a detailed explanation of how the rate data was calculated and used for each of the reactions in \autoref{sec:reaction_list}.

\subsection{Collisional electronic excitation from $X^1\Sigma_g$}
\label{sec:Coll_ex_X}
The cross sections for electronic collisional excitation from $X^1\Sigma_g$ are obtained from the MCCCDB website \cite{MCCCDB_exc_X1}. These cross sections are given specifically for D$_2$. The cross sections are only given for excitation from the ground state to higher electronic states, so collisional deexcitation cross sections to the ground state are calculated using the principle of detailed balance, explained in \ref{sec:det_balance}. To obtain rate coefficients, the cross sections are extrapolated linearly in loglog space as a function of energy up to 5000 eV. Next, \eqref{eq:Bolzmann_integral} is used to obtain rate coefficients as a function of temperature. The rate coefficients are then fitted using a logarithmic polynomial fit of the form 
\begin{equation}
    \ln(\langle\sigma v\rangle_{eff}) \approx b_0 + b_1 \ln(T) + b_2 (\ln(T))^2 + ... + b_8 (\ln(T))^8
\end{equation}
This fit is valid for a temperature range of 0.1 to 100 eV. This format is similar to how Eirene stores 1D effective rate data, which facilitates using the fit coefficients $b_0$ to $b_8$ as input for CRUMPET to construct the rate matrices. 

\subsection{Molecular charge exchange}
The vibrationally resolved rate coefficients for molecular charge exchange are obtained from Ichihara et al. for molecular hydrogen \cite{Ichihara_2000}. The isotope mass difference for deuterium is accounted for by using $T_i=T_e/2$ for the ion temperature \cite{Greenland2001,Kotov2007}. The rate coefficients are only given for $T<5$ eV, so for higher temperatures (up to 100 eV), they were extrapolated linearly in loglog space. The rate coefficients were then fitted in a similar manner to \ref{sec:Coll_ex_X} and used as input data for CRUMPET.

\subsection{Vibrational transitions in $X^1\Sigma_g$ and $B^1\Sigma_g$ through electron-impact}
The cross sections for vibrational transitions in $X^1\Sigma_g$ and $B^1\Sigma_g$ through electron-impact are obtained from the LXCat website (www.lxcat.net), specifically from the Laporta database \cite{Laporta_2021}. These cross sections were given specifically for $D_2$. Again the principle of detailed balance was used for transitions from high to low vibrational quantum numbers. The cross sections were extrapolated, converted to rates and fitted in a similar way to \ref{sec:Coll_ex_X} as function of electron temperature. As opposed to the Eirene 'H2VIBR' rates, this considers reactions from any vibrational level to any other vibrational level, not just $\nu \+- 1$

\subsection{Dissociative excitation from $X^1\Sigma_g$}
The cross sections for dissociative excitation from $X^1\Sigma_u$ are obtained from MCCCDB \cite{MCCCDB_exc_X1}. The cross sections were calculated by adding the cross sections together for excitation from the ground state to all the repulsive higher states in the MCCCDB database. The cross sections were calculated specifically for D$_2$, and were extrapolated, converted to rate constants and fitted in the same way as \ref{sec:Coll_ex_X}. 

\subsection{Dissociative attachment}
The cross sections for dissociative attachment were obtained from the Laporta database via the LXCat website \cite{Laporta_2021}. These cross sections show many resonances, which makes them difficult to extrapolate outside of the energy domain for which they are explicitly calculated. Therefore, the cross sections are not extrapolated, and are assumed to be 0 outside of the given domain. The conversion to rate coefficients and fitting is done in the same way as \ref{sec:Coll_ex_X}. The cross sections for dissociative attachment were calculated specifically for D$_2$.

\subsection{$D_2$ ionization}
The $D_2$ ionization cross sections are obtained from the MCCCDB website \cite{MCCCDB_Ionization}. The $D_2$ ionization from $d^3\Pi_u$ state is not given in MCCCDB, so its cross section was assumed to be equal to that of $c^3\Pi_u$. The extrapolation, rate constant calculation and fitting was again done in the same way as in \ref{sec:Coll_ex_X}. These cross sections were also calculated specifically for D$_2$.

\subsection{Collisional excitation between higher states.}
The cross sections collisional excitation from electronically excited states are not available in the MCCCDB or Laporta database for deuterium. The MCCCDB does contain excitation cross sections for hydrogen \cite{MCCCDB_exc_exc}. These cross sections are only given as a sum over all the vibrational levels in the upper state:
\begin{equation}
    e+\text{D}_2(n=A,\nu=0)\xrightarrow{\langle\sigma v\rangle_i} e+\text{D}_2(n=B)
\end{equation}
Since our model contains vibrationally resolved excited states, we actually need the cross sections for the reaction
\begin{equation}
    e+\text{D}_2(n=A,\nu=i)\xrightarrow{\langle\sigma v\rangle_{ij}} e+\text{D}_2(n=B,\nu=j)
\end{equation}
To calculate the cross sections $\sigma_{ij}$ from the available $\sigma_0$, we use the Franck-Condon factors from Fantz et al. \cite{FANTZ2006853}. 
\begin{equation}
    \sigma_{ij} = \frac{\sigma_0K_{ij}}{\sum_{k=0}^{14}K_{0k}}
\end{equation}
The cross sections for deexcitation are calculated using the principle of detailed balance. The fit constants for the rate constants were obtained in a similar way to \ref{sec:Coll_ex_X}.
The Einstein coefficients for radiative deexcitation are taken from Fantz et al. \cite{FANTZ2006853}.

\subsection{Reactions with ions}
Rates that consume D$_2^+$ and D$^-$ are taken from the AMJUEL documentation of Eirene \cite{Eirene}, and are therefore only given for hydrogen, not deuterium. 

\subsection{Vibrational-Vibrational Exchange (VVE)}
\label{sec:VVE_rate}
The rate constants used for VVE are calculated using a formula from \cite{Matveyev_1995}.

\begin{equation}
    k^{v,v+1}_{w+1,w} \approx (v+1)(w+1)k^{0,1}_{1,0}\left(\frac{3}{2}-\frac{1}{2}e^{-\delta(v-w)}\right)\exp(\Delta_1(v-w)-\Delta_2(v-w)^2),
\end{equation}
with $k^{0,1}_{1,0}\approx4.23\cdot10^{-15}(300/T)^{1/3}$ cm$^3$s$^{-1}$, $\delta\approx 0.21(T/300)^{1/2}$, $\Delta_1\approx0.236(T/300)^{1/4}$, and $\Delta_2 \approx 0.0572(T/300)^{1/2}$. Here $T$ is the temperature of the D$_2$ molecules in Kelvin. The rate for VVE was calculated in \cite{Matveyev_1995} for hydrogen, not deuterium. 

%% file: Appendix_C.tex
\section{Scaling laws for molecular and atomic density inclusion}
\label{sec:scalings}
In sections \ref{sec:rates} and \ref{sec:radiative_loss}, scaling laws were used for the atomic and molecular densities to investigate the relative importance of atomic and molecular processes. The scaling laws were taken from a paper by Verhaegh et al. from 2023, who generated them for the MAST-U Super-X divertor \cite{Verhaegh_2023_fdens}. They are given here below: 

\begin{align}
    f_\text{D}(T_e)&=10^{-1.4721-1.126 \log10(T_e)} \text{m} \\
    f_{\text{D}_2}(T_e)&=10^{-1.176-2.3075 \log10(T_e)} \text{m} \\
    f_{D^+}(T_e)&=10^{-0.2609} \text{m}
\end{align}
where $T_e$ is the electron temperature in eV.

%% file: bibliography.bib
@article{Verhaegh_2021,
doi = {10.1088/1741-4326/ac1dc5},
url = {https://dx.doi.org/10.1088/1741-4326/ac1dc5},
year = {2021},
month = {sep},
publisher = {IOP Publishing},
volume = {61},
number = {10},
pages = {106014},
author = {K. Verhaegh and B. Lipschultz and J.R. Harrison and B.P. Duval and A. Fil and M. Wensing and C. Bowman and D.S. Gahle and A. Kukushkin and D. Moulton and A. Perek and A. Pshenov and F. Federici and O. Février and O. Myatra and A. Smolders and C. Theiler and  the TCV Team and  the EUROfusion MST1 Team},
title = {The role of plasma-molecule interactions on power and particle balance during detachment on the TCV tokamak},
journal = {Nuclear Fusion}
}

@article{Greenland_Molecules,
 ISSN = {13645021},
 URL = {http://www.jstor.org/stable/3067378},
 author = {P. T. Greenland},
 journal = {Proceedings: Mathematical, Physical and Engineering Sciences},
 number = {2012},
 pages = {1821--1839},
 publisher = {The Royal Society},
 title = {Collisional-Radiative Models with Molecules},
 urldate = {2023-05-24},
 volume = {457},
 year = {2001}
}

@article{Myatra2023,
doi = {10.1088/1741-4326/acd9da},
url = {https://dx.doi.org/10.1088/1741-4326/acd9da},
year = {2023},
month = {jun},
publisher = {IOP Publishing},
volume = {63},
number = {7},
pages = {076030},
author = {O. Myatra and D. Moulton and B. Dudson and B. Lipschultz and S. Newton and K. Verhaegh and A. Fil},
title = {SOLPS-ITER predictive simulations of the impact of ion-molecule elastic collisions on strongly detached MAST-U Super-X divertor conditions},
journal = {Nuclear Fusion}
}

@Article{Sawada2016,
AUTHOR = {Sawada, Keiji and Goto, Motoshi},
TITLE = {Rovibrationally Resolved Time-Dependent Collisional-Radiative Model of Molecular Hydrogen and Its Application to a Fusion Detached Plasma},
JOURNAL = {Atoms},
VOLUME = {4},
YEAR = {2016},
NUMBER = {4},
ARTICLE-NUMBER = {29},
URL = {https://www.mdpi.com/2218-2004/4/4/29},
ISSN = {2218-2004},
DOI = {10.3390/atoms4040029}
}

@TechReport{Kotov2007,
  author        = {Kotov, Vladislav and Reiter, Detlev and Kukushkin, Andrey S},
  title         = {Numerical study of the ITER divertor plasma with the B2-EIRENE code package},
  year          = {2007},
  __markedentry = {[kevin:6]},
  type          = {J\"{u}lich report - JUEL 4257},
  url           = {http://www.eirene.de/kotov\_solps42\_report.pdf},
  institution   = {Forschungszentrum Jülich GmbH}
}

@article{Vankan2004,
author = {Vankan,P.  and Heil,S. B. S.  and Mazouffre,S.  and Engeln,R.  and Schram,D. C.  and Döbele,H. F. },
title = {{A vacuum-UV laser-induced fluorescence experiment for measurement of rotationally and vibrationally excited H2}},
journal = {Review of Scientific Instruments},
volume = {75},
number = {4},
pages = {996-999},
year = {2004},
doi = {10.1063/1.1688435},
}

@PhdThesis{Wischmeier2005,
  author        = {Wischmeier, Marco},
  title         = {Simulating divertor detachment in the TCV and JET tokamaks},
  year          = {2005},
  type          = {Thesis},
  __markedentry = {[kevin:6]},
  doi           = {10.5075/epfl-thesis-3176},
  school    = {EPFL},
}

@Article{Krasheninnikov1997,
  author        = {Krasheninnikov, SI and Pigarov, A Yu and Knoll, DA and LaBombard, B and Lipschultz, B and Sigmar, DJ and Soboleva, TK and Terry, JL and Wising, F},
  title         = {Plasma recombination and molecular effects in tokamak divertors and divertor simulators},
  journal       = {Physics of Plasmas},
  year          = {1997},
  volume        = {4},
  number        = {5},
  pages         = {1638-1646},
  issn          = {1070-664X},
  __markedentry = {[kevin:6]},
  type          = {Journal Article},
  doi           = {10.1063/1.872268},
}

@TechReport{Reiter2018,
title = {Isotope effects in molecule assisted recombination and dissociation in divertor plasmas},
series = {Berichte des Forschungszentrums Jülich ;},
author = {Janev, R. K. and Reiter, D.},
address = {Jülich},
publisher = {Forschungszentrum, Zentralbibliothek},
year = {2018},
pages = {1 Online-Ressource (37 Seiten)},
note = {englisch},
type = {J\"{u}lich report - JUEL 4411},
url = {https://juser.fz-juelich.de/record/850290/files/J\%C3\%BCl\_4411\_Reiter.pdf?version=1},
institution   = {Forschungszentrum Jülich GmbH}
}

@article{Krishnakumar2011,
  title = {Dissociative Electron Attachment Cross Sections for ${\mathrm{H}}_{2}$ and ${\mathrm{D}}_{2}$},
  author = {Krishnakumar, E. and Denifl, S. and \ifmmode \check{C}\else \v{C}\fi{}ade\ifmmode \check{z}\else \v{z}\fi{}, I. and Markelj, S. and Mason, N. J.},
  journal = {Phys. Rev. Lett.},
  volume = {106},
  issue = {24},
  pages = {243201},
  numpages = {4},
  year = {2011},
  month = {Jun},
  publisher = {American Physical Society},
  doi = {10.1103/PhysRevLett.106.243201},
}

@Article{Stangeby2017,
  author        = {Stangeby, P. C. and Chaofeng, Sang},
  title         = {Strong correlation between {$D_2$} density and electron temperature at the target of divertors found in SOLPS analysis},
  journal       = {Nuclear Fusion},
  year          = {2017},
  volume        = {57},
  number        = {5},
  pages         = {056007},
  issn          = {0029-5515},
  __markedentry = {[kevin:6]},
  type          = {Journal Article},
  doi           = {10.1088/1741-4326/aa5e27},
}

@techreport{Takayanagi1977,
  title={Cross sections for atomic processes, vol. 2},
  author={Takayanagi, Kazuo and Suzuki, Hiroshi and Otani, Shunsuke},
  year={1977},
  institution={Nagoya Univ.(Japan). Inst. of Plasma Physics}
}

@book{Janev1987,
  title={Elementary processes in hydrogen-helium plasmas: cross sections and reaction rate coefficients},
  author={Janev, Ratko K and Langer, William D and Douglass Jr, E and others},
  year={1987},
  publisher={Springer Science \& Business Media}
}

@article{Sawada1995,
  title={Effective ionization and dissociation rate coefficients of molecular hydrogen in plasma},
  author={Sawada, Keiji and Fujimoto, Takashi},
  journal={Journal of applied physics},
  volume={78},
  number={5},
  pages={2913--2924},
  year={1995},
  publisher={American Institute of Physics}
}

@article{Holliday1971,
author = {Holliday,M. G.  and Muckerman,J. T.  and Friedman,L. },
title = {Isotopic Studies of the Proton–Hydrogen Molecule Reaction},
journal = {The Journal of Chemical Physics},
volume = {54},
number = {3},
pages = {1058-1072},
year = {1971},
doi = {10.1063/1.1674939},
}

@Article{Fantz2002,
  author        = {Fantz, U.},
  title         = {Emission spectroscopy of hydrogen molecules in technical and divertor plasmas},
  journal       = {Contributions to Plasma Physics},
  year          = {2002},
  volume        = {42},
  number        = {6-7},
  pages         = {675-684},
  issn          = {0863-1042},
  __markedentry = {[kevin:6]},
  type          = {Journal Article},
}

@Article{Kukushkin2017,
  author        = {Kukushkin, A. S. and Krasheninnikov, S. I. and Pshenov, A. A. and Reiter, D.},
  title         = {Role of molecular effects in divertor plasma recombination},
  journal       = {Nuclear Materials and Energy},
  year          = {2017},
  volume        = {12},
  pages         = {984-988},
  issn          = {2352-1791},
  __markedentry = {[kevin:6]},
  doi           = {10.1016/j.nme.2016.12.030},
  type          = {Journal Article},
}

@TechReport{Greenland2001,
title = {The CRMOL manual: collisional-radiative models for molecular hydrogen in plasmas},
author = {Greenland, P T},
place = {Germany},
year = {2001},
month = {Apr},
type = {J\"{u}lich report Juel-3858},
url   = {https://juser.fz-juelich.de/record/24992/files/J\%C3\%BCl\_3858\_Greenland.pdf?version=1},
institution   = {Forschungszentrum Jülich GmbH}
}

@article{FANTZ2006853,
title = {Franck–Condon factors, transition probabilities, and radiative lifetimes for hydrogen molecules and their isotopomeres},
journal = {Atomic Data and Nuclear Data Tables},
volume = {92},
number = {6},
pages = {853-973},
year = {2006},
issn = {0092-640X},
doi = {https://doi.org/10.1016/j.adt.2006.05.001},
url = {https://www.sciencedirect.com/science/article/pii/S0092640X06000386},
author = {U. Fantz and D. Wünderlich}
}

@article{article,
author = {Gough, S. and Schermann, C. and Pichou, F. and Landau, M. and Čadež, Iztok and Hall, R.},
year = {1995},
month = {12},
pages = {687},
title = {The formation of vibrationally excited hydrogen molecules on carbon surfaces.},
volume = {305},
journal = {Astronomy and Astrophysics}
}

@article{MCCCDB_Ionization,
title = {Complete collision data set for electrons scattering on molecular hydrogen and its isotopologues: IV. Vibrationally-resolved ionization of the ground and excited electronic states},
journal = {Atomic Data and Nuclear Data Tables},
volume = {151},
pages = {101573},
year = {2023},
issn = {0092-640X},
doi = {https://doi.org/10.1016/j.adt.2023.101573},
url = {https://www.sciencedirect.com/science/article/pii/S0092640X23000013},
author = {Liam H. Scarlett and Eric Jong and Starsha Odelia and Mark C. Zammit and Yuri Ralchenko and Barry I. Schneider and Igor Bray and Dmitry V. Fursa}
}

@article{Ichihara_2000,
doi = {10.1088/0953-4075/33/21/318},
url = {https://dx.doi.org/10.1088/0953-4075/33/21/318},
year = {2000},
month = {nov},
publisher = {},
volume = {33},
number = {21},
pages = {4747},
author = {Akira Ichihara and  Osamu Iwamoto and  R K Janev},
title = {Cross sections for the
reaction
H+ + H2 (v = 0-14) →H + H2+ at low collision
energies},
journal = {Journal of Physics B: Atomic, Molecular and Optical Physics}
}

@article{MCCCDB_exc_X1,
title = {Complete collision data set for electrons scattering on molecular hydrogen and its isotopologues: I. Fully vibrationally-resolved electronic excitation of H2(X1g+)},
journal = {Atomic Data and Nuclear Data Tables},
volume = {137},
pages = {101361},
year = {2021},
issn = {0092-640X},
doi = {https://doi.org/10.1016/j.adt.2020.101361},
url = {https://www.sciencedirect.com/science/article/pii/S0092640X20300292},
author = {Liam H. Scarlett and Dmitry V. Fursa and Mark C. Zammit and Igor Bray and Yuri Ralchenko and Kayla D. Davie}
}

@article{MCCCDB_exc_exc,
  title = {Convergent close-coupling calculations of electrons scattering on electronically excited molecular hydrogen},
  author = {Scarlett, Liam H. and Savage, Jeremy S. and Fursa, Dmitry V. and Bray, Igor and Zammit, Mark C. and Schneider, Barry I.},
  journal = {Phys. Rev. A},
  volume = {103},
  issue = {3},
  pages = {032802},
  numpages = {17},
  year = {2021},
  month = {Mar},
  publisher = {American Physical Society},
  doi = {10.1103/PhysRevA.103.032802},
  url = {https://link.aps.org/doi/10.1103/PhysRevA.103.032802}
}

@article{Eirene,
author = {D. Reiter and M. Baelmans and P. Börner},
title = {The EIRENE and B2-EIRENE Codes},
journal = {Fusion Science and Technology},
volume = {47},
number = {2},
pages = {172-186},
year  = {2005},
publisher = {Taylor & Francis},
doi = {10.13182/FST47-172},

URL = { 
    
        https://doi.org/10.13182/FST47-172
    
    

},
eprint = { 
    
        https://doi.org/10.13182/FST47-172
    
    

}

}

@article{Laporta_2021,
doi = {10.1088/1361-6587/ac0163},
url = {https://dx.doi.org/10.1088/1361-6587/ac0163},
year = {2021},
month = {jun},
publisher = {IOP Publishing},
volume = {63},
number = {8},
pages = {085006},
author = {V Laporta and R Agnello and G Fubiani and I Furno and C Hill and D Reiter and F Taccogna},
title = {Vibrational excitation and dissociation of deuterium molecule by electron impact},
journal = {Plasma Physics and Controlled Fusion}
}

@article{Verhaegh_2023_fdens,
doi = {10.1088/1741-4326/aca10a},
url = {https://dx.doi.org/10.1088/1741-4326/aca10a},
year = {2022},
month = {dec},
publisher = {IOP Publishing},
volume = {63},
number = {1},
pages = {016014},
author = {K. Verhaegh and B. Lipschultz and J.R. Harrison and N. Osborne and A.C. Williams and P. Ryan and J. Allcock and J.G. Clark and F. Federici and B. Kool and T. Wijkamp and A. Fil and D. Moulton and O. Myatra and A. Thornton and T.O.S.J. Bosman and C. Bowman and G. Cunningham and B.P. Duval and S. Henderson and R. Scannell and  the MAST Upgrade team},
title = {Spectroscopic investigations of detachment on the MAST Upgrade Super-X divertor},
journal = {Nuclear Fusion}
}

@article{Verhaegh_2023_SOLPS,
doi = {10.1088/1741-4326/acd394},
url = {https://dx.doi.org/10.1088/1741-4326/acd394},
year = {2023},
month = {may},
publisher = {IOP Publishing},
volume = {63},
number = {7},
pages = {076015},
author = {K. Verhaegh and A.C. Williams and D. Moulton and B. Lipschultz and B.P. Duval and O. Février and A. Fil and J. Harrison and N. Osborne and H. Reimerdes and C. Theiler and  the TCV Team and the EUROfusion MST1 Team},
title = {Investigating the impact of the molecular charge-exchange rate on detached SOLPS-ITER simulations},
journal = {Nuclear Fusion}
}

@article{Rutigliano_Cacciatore_2011, title={Eley–rideal recombination of hydrogen atoms on a tungsten surface}, volume={13}, DOI={10.1039/c0cp02514c}, number={16}, journal={Physical Chemistry Chemical Physics}, author={Rutigliano, M. and Cacciatore, M.}, year={2011}, pages={7475}}

@misc{ITER_2023, url={https://www.iter.org/mach/Divertor}, journal={ITER}, year={2023}}

@article{Fil_2022,
doi = {10.1088/1741-4326/ac81d8},
url = {https://dx.doi.org/10.1088/1741-4326/ac81d8},
year = {2022},
month = {aug},
publisher = {IOP Publishing},
volume = {62},
number = {9},
pages = {096026},
author = {A. Fil and B. Lipschultz and D. Moulton and A. Thornton and B.D. Dudson and O. Myatra and K. Verhaegh and  the EUROfusion MST1 Team},
title = {Comparison between MAST-U conventional and Super-X configurations through SOLPS-ITER modelling},
journal = {Nuclear Fusion}
}

@misc{EUROfusion_2023, url={https://euro-fusion.org/programme/demo/}, journal={EUROfusion}, year={2023}, month={Jan}}

@article{Reimerdes_2020,
doi = {10.1088/1741-4326/ab8a6a},
url = {https://dx.doi.org/10.1088/1741-4326/ab8a6a},
year = {2020},
month = {may},
publisher = {IOP Publishing},
volume = {60},
number = {6},
pages = {066030},
author = {H. Reimerdes and R. Ambrosino and P. Innocente and A. Castaldo and P. Chmielewski and G. Di Gironimo and S. Merriman and V. Pericoli-Ridolfini and L. Aho-Mantilla and R. Albanese and H. Bufferand and G. Calabro and G. Ciraolo and D. Coster and N. Fedorczak and S. Ha and R. Kembleton and K. Lackner and V.P. Loschiavo and T. Lunt and D. Marzullo and R. Maurizio and F. Militello and G. Ramogida and F. Subba and S. Varoutis and R. Zagórski and H. Zohm},
title = {Assessment of alternative divertor configurations as an exhaust solution for DEMO},
journal = {Nuclear Fusion}
}

@article{Matveyev_1995,
doi = {10.1088/0963-0252/4/4/012},
url = {https://dx.doi.org/10.1088/0963-0252/4/4/012},
year = {1995},
month = {nov},
publisher = {},
volume = {4},
number = {4},
pages = {606},
author = {A A Matveyev and  V P Silakov},
title = {Kinetic processes in a highly-ionized non-equilibrium hydrogen plasma},
journal = {Plasma Sources Science and Technology}
}

@article{HOLM2021100982,
title = {Comparison of a collisional-radiative fluid model of H2 in UEDGE to the kinetic neutral code EIRENE},
journal = {Nuclear Materials and Energy},
volume = {27},
pages = {100982},
year = {2021},
issn = {2352-1791},
doi = {https://doi.org/10.1016/j.nme.2021.100982},
url = {https://www.sciencedirect.com/science/article/pii/S235217912100065X},
author = {A. Holm and P. Börner and T.D. Rognlien and W.H Meyer and M. Groth}
}

@misc{Yacora_on_the_Web, url={https://www.yacora.de/help/help-page-for-h2-model}, journal={Yacora on the Web}, publisher={Max Planck Institute for Plasma Physics}}

@misc{verhaegh2023role,
      title={The role of plasma-atom and molecule interactions on power \& particle balance during detachment on the MAST Upgrade Super-X divertor}, 
      author={Kevin Verhaegh and Bruce Lipschultz and James Harrison and Fabio Federici and David Moulton and Nicola Lonigro and Stijn Kobussen and Martin O'Mullane and Nick Osborne and Peter Ryan and Tijs Wijkamp and Bob Kool and Effy Rose and Christian Theiler and Andrew Thornton},
      year={2023},
      eprint={2304.09109},
      archivePrefix={arXiv},
      primaryClass={physics.plasm-ph}
}

@article{WIESEN2015480,
title = {The new SOLPS-ITER code package},
journal = {Journal of Nuclear Materials},
volume = {463},
pages = {480-484},
year = {2015},
note = {PLASMA-SURFACE INTERACTIONS 21},
issn = {0022-3115},
doi = {https://doi.org/10.1016/j.jnucmat.2014.10.012},
url = {https://www.sciencedirect.com/science/article/pii/S0022311514006965},
author = {S. Wiesen and D. Reiter and V. Kotov and M. Baelmans and W. Dekeyser and A.S. Kukushkin and S.W. Lisgo and R.A. Pitts and V. Rozhansky and G. Saibene and I. Veselova and S. Voskoboynikov}
}

@misc{osborne2023initial,
      title={Initial Fulcher band observations from high resolution spectroscopy in the MAST-U divertor}, 
      author={N. Osborne and K. Verhaegh and M. D. Bowden and T. Wijkamp and N. Lonigro and P. Ryan and B. Lipschultz and V. Soukhanovskii and T. van den Biggelaar and the MAST-U team},
      year={2023},
      eprint={2306.16969},
      archivePrefix={arXiv},
      primaryClass={physics.plasm-ph}
}
